\documentclass[twocolumn]{aastex631}

\usepackage{amssymb,amsmath,comment,bm}
\usepackage{graphicx}
\usepackage{booktabs}

\shorttitle{Methods of Error Estimation for $21\,\textrm{cm}$ Delay Power Spectra}
\shortauthors{Tan et al.}

\graphicspath{{./}{figures/}}

\begin{document}

\title{Methods of Error Estimation for Delay Power Spectra in $21\,\textrm{cm}$ Cosmology}

\correspondingauthor{Jianrong Tan}
\email{jianrong@sas.upenn.edu}

\author[0000-0001-6161-7037]{Jianrong Tan}
\affiliation{Department of Physics and Astronomy, University of Pennsylvania, Philadelphia, PA 19104, USA}
\affiliation{Department of Physics and McGill Space Institute, McGill University, Montreal, QC, Canada H3A 2T8}

\author[0000-0001-6876-0928]{Adrian Liu}
\affiliation{Department of Physics and McGill Space Institute, McGill University, Montreal, QC, Canada H3A 2T8}

\author[0000-0002-8211-1892]{Nicholas S. Kern}
\affiliation{Department of Physics, Massachusetts Institute of Technology, Cambridge, MA, USA}

\author{Zara  Abdurashidova}
\affiliation{Department of Astronomy, University of California, Berkeley, CA}

\author{James E. Aguirre}
\affiliation{Department of Physics and Astronomy, University of Pennsylvania, Philadelphia, PA 19104, USA}

\author{Paul  Alexander}
\affiliation{Cavendish Astrophysics, University of Cambridge, Cambridge, UK}

\author{Zaki S. Ali}
\affiliation{Department of Astronomy, University of California, Berkeley, CA}

\author{Yanga  Balfour}
\affiliation{SKA-SA, Cape Town, South Africa}

\author{Adam P. Beardsley}
\affiliation{School of Earth and Space Exploration, Arizona State University, Tempe, AZ}

\author{Gianni  Bernardi}
\affiliation{Department of Physics and Electronics, Rhodes University, PO Box 94, Grahamstown, 6140, South Africa}
\affiliation{INAF-Istituto di Radioastronomia, via Gobetti 101, 40129 Bologna, Italy}
\affiliation{SKA-SA, Cape Town, South Africa}

\author{Tashalee S. Billings}
\affiliation{Department of Physics and Astronomy, University of Pennsylvania, Philadelphia, PA 19104, USA}

\author{Judd D. Bowman}
\affiliation{School of Earth and Space Exploration, Arizona State University, Tempe, AZ}

\author{Richard F. Bradley}
\affiliation{National Radio Astronomy Observatory, Charlottesville, VA}

\author{Philip Bull}
\affiliation{School of Physics \& Astronomy, Queen Mary University of London, London, UK}

\author{Jacob  Burba}
\affiliation{Department of Physics, Brown University, Providence, RI}

\author{Steven  Carey}
\affiliation{Cavendish Astrophysics, University of Cambridge, Cambridge, UK}

\author{Christopher L. Carilli}
\affiliation{National Radio Astronomy Observatory, Socorro, NM}

\author{Carina  Cheng}
\affiliation{Department of Astronomy, University of California, Berkeley, CA}

\author{David R. DeBoer}
\affiliation{Department of Astronomy, University of California, Berkeley, CA}

\author{Matt  Dexter}
\affiliation{Department of Astronomy, University of California, Berkeley, CA}

\author{Eloy  de~Lera~Acedo}
\affiliation{Cavendish Astrophysics, University of Cambridge, Cambridge, UK}

\author[0000-0003-3336-9958]{Joshua S. Dillon}
\affiliation{Department of Astronomy, University of California, Berkeley, CA}
%\affiliation{NSF Astronomy and Astrophysics Postdoctoral Fellow}

\author{John  Ely}
\affiliation{Cavendish Astrophysics, University of Cambridge, Cambridge, UK}

\author{Aaron Ewall-Wice}
\affiliation{Department of Astronomy, University of California, Berkeley, CA}

\author{Nicolas  Fagnoni}
\affiliation{Cavendish Astrophysics, University of Cambridge, Cambridge, UK}

\author{Randall  Fritz}
\affiliation{SKA-SA, Cape Town, South Africa}

\author{Steve R. Furlanetto}
\affiliation{Department of Physics and Astronomy, University of California, Los Angeles, CA}

\author{Kingsley  Gale-Sides}
\affiliation{Cavendish Astrophysics, University of Cambridge, Cambridge, UK}

\author{Brian  Glendenning}
\affiliation{National Radio Astronomy Observatory, Socorro, NM}

\author{Deepthi  Gorthi}
\affiliation{Department of Astronomy, University of California, Berkeley, CA}

\author{Bradley  Greig}
\affiliation{School of Physics, University of Melbourne, Parkville, VIC 3010, Australia}

\author{Jasper  Grobbelaar}
\affiliation{SKA-SA, Cape Town, South Africa}

\author{Ziyaad  Halday}
\affiliation{SKA-SA, Cape Town, South Africa}

\author{Bryna J. Hazelton}
\affiliation{Department of Physics, University of Washington, Seattle, WA}
\affiliation{eScience Institute, University of Washington, Seattle, WA}

\author{Jacqueline N. Hewitt}
\affiliation{Department of Physics, Massachusetts Institute of Technology, Cambridge, MA}

\author{Jack  Hickish}
\affiliation{Department of Astronomy, University of California, Berkeley, CA}

\author{Daniel C. Jacobs}
\affiliation{School of Earth and Space Exploration, Arizona State University, Tempe, AZ}

\author{Austin  Julius}
\affiliation{SKA-SA, Cape Town, South Africa}

\author{Joshua  Kerrigan}
\affiliation{Department of Physics, Brown University, Providence, RI}

\author{Piyanat  Kittiwisit}
\affiliation{School of Chemistry and Physics, University of KwaZulu-Natal, Westville Campus, Durban, South Africa}

\author{Saul A. Kohn}
\affiliation{Department of Physics and Astronomy, University of Pennsylvania, Philadelphia, PA 19104, USA}

\author{Matthew  Kolopanis}
\affiliation{School of Earth and Space Exploration, Arizona State University, Tempe, AZ}

\author{Adam  Lanman}
\affiliation{Department of Physics, Brown University, Providence, RI}

\author{Paul  La~Plante}
\affiliation{Department of Astronomy, University of California, Berkeley, CA}

\author{Telalo  Lekalake}
\affiliation{SKA-SA, Cape Town, South Africa}

\author{David  MacMahon}
\affiliation{Department of Astronomy, University of California, Berkeley, CA}

\author{Lourence  Malan}
\affiliation{SKA-SA, Cape Town, South Africa}

\author{Cresshim  Malgas}
\affiliation{SKA-SA, Cape Town, South Africa}

\author{Matthys  Maree}
\affiliation{SKA-SA, Cape Town, South Africa}

\author{Zachary E. Martinot}
\affiliation{Department of Physics and Astronomy, University of Pennsylvania, Philadelphia, PA 19104, USA}

\author{Eunice  Matsetela}
\affiliation{SKA-SA, Cape Town, South Africa}

\author{Andrei  Mesinger}
\affiliation{Scuola Normale Superiore, 56126 Pisa, PI, Italy}

\author{Mathakane  Molewa}
\affiliation{SKA-SA, Cape Town, South Africa}

\author{Miguel F. Morales}
\affiliation{Department of Physics, University of Washington, Seattle, WA}

\author{Tshegofalang  Mosiane}
\affiliation{SKA-SA, Cape Town, South Africa}

\author{Steven G. Murray}
\affiliation{School of Earth and Space Exploration, Arizona State University, Tempe, AZ}

\author{Abraham R. Neben}
\affiliation{Department of Physics, Massachusetts Institute of Technology, Cambridge, MA, USA}

\author{Bojan  Nikolic}
\affiliation{Cavendish Astrophysics, University of Cambridge, Cambridge, UK}

\author{Chuneeta D. Nunhokee}
\affiliation{Department of Astronomy, University of California, Berkeley, CA}

\author{Aaron R. Parsons}
\affiliation{Department of Astronomy, University of California, Berkeley, CA}

\author{Nipanjana  Patra}
\affiliation{Department of Astronomy, University of California, Berkeley, CA}

\author{Samantha  Pieterse}
\affiliation{SKA-SA, Cape Town, South Africa}

\author{Jonathan C. Pober}
\affiliation{Department of Physics, Brown University, Providence, RI}

\author{Nima  Razavi-Ghods}
\affiliation{Cavendish Astrophysics, University of Cambridge, Cambridge, UK}

\author{Jon  Ringuette}
\affiliation{Department of Physics, University of Washington, Seattle, WA}

\author{James  Robnett}
\affiliation{National Radio Astronomy Observatory, Socorro, NM}

\author{Kathryn  Rosie}
\affiliation{SKA-SA, Cape Town, South Africa}

\author{Peter  Sims}
\affiliation{Department of Physics, Brown University, Providence, RI}

\author{Saurabh Singh}
\affiliation{Department of Physics and McGill Space Institute, McGill University, Montreal, QC, Canada H3A 2T8}

\author{Craig  Smith}
\affiliation{SKA-SA, Cape Town, South Africa}

\author{Angelo  Syce}
\affiliation{SKA-SA, Cape Town, South Africa}

\author{Nithyanandan  Thyagarajan}
\affiliation{School of Earth and Space Exploration, Arizona State University, Tempe, AZ}
\affiliation{National Radio Astronomy Observatory, Socorro, NM}

\author{Peter K.~G. Williams}
\affiliation{Center for Astrophysics, Harvard \& Smithsonian, 60 Garden St., Cambridge, MA}
\affiliation{American Astronomical Society, 1667 K Street NW, Suite 800, Washington, DC 20006}

\author{Haoxuan  Zheng}
\affiliation{Department of Physics, Massachusetts Institute of Technology, Cambridge, MA}

\begin{abstract}
Precise measurements of the 21 cm power spectrum are crucial for understanding the physical processes of hydrogen reionization. Currently, this probe is being pursued by low-frequency radio interferometer arrays. As these experiments come closer to making a first detection of the signal, error estimation will play an increasingly important role in setting robust measurements. Using the delay power spectrum approach, we have produced a critical examination of different ways that one can estimate error bars on the power spectrum. We do this through a synthesis of analytic work, simulations of toy models, and tests on small amounts of real data. We find that, although computed independently, the different error bar methodologies are in good agreement with each other in the noise-dominated regime of the power spectrum. For our preferred methodology, the predicted probability distribution function is consistent with the empirical noise power distributions from both simulated and real data. This diagnosis is mainly in support of the forthcoming HERA upper limit, and also is expected to be more generally applicable.
\end{abstract}

\section{Introduction} 
\label{sec:intro}
%%%
The Epoch of Reionization (EoR)---when neutral hydrogen in the intergalactic medium (IGM) was ionized by photons from early galaxies and active galactic nuclei---remains one of the most exciting frontiers in modern astrophysics and cosmology. Precise measurements of this era will significantly enhance our understanding on the origin of very first stars, the process of galaxy formation and the thermal history of the IGM \citep{barkana2001beginning,dayal2018early}. Some measurements, such as those of the optical depth of Cosmic Microwave Background (CMB) photons \citep{aghanim2018planck}, the Gunn-Peterson trough in distant quasar spectra \citep{becker2001evidence, fan2006observational,bolton2011neutral,becker2015evidence}, quasar damping wings \citep{davies2018quantitative}, and the decrease in the number density and the clustering trends of  Ly-$\alpha$ emitters at high redshifts \citep{stark2010keck, ouchi2010statistics, bosman2018new}, have already established the basic parameters of the EoR. Collectively, they suggest that reionization is a process which probably began at $z \gg 10$ and ended around $z \approx 6$. However, the aforementioned probes paint an indirect and incomplete picture of the EoR. For example, CMB measurements are integral constraints over redshift, making the extraction of detailed information technically difficult (often involving subtle kinetic Sunyaev-Zel'dovich effect or polarization measurements); Ly$\alpha$ photons suffer from severely saturated absorption that makes it difficult for them to probe earlier times than the end of reionization; and low-mass galaxies (i.e., those thought to be responsible for supplying a large fraction of ionizing photons) are too faint to be directly detected. A complementary probe capable of making direct observations of the EoR is therefore desirable.

A strong candidate for a direct probe of reionization is the $21\,\textrm{cm}$ line. Arising from the ``spin flip'' transition in the hyperfine structure of atomic hydrogen, the $21\,\textrm{cm}$ line is a promising way to directly trace the evolution of HI regimes on different spatial scales and to eventually provide a comprehensive three-dimensional picture throughout the history of reionization \citep{furlanetto2006cosmology, morales2010reionization, pritchard201221, liu2019data}. Current experimental efforts are focused on slightly more modest---but still ambitious---observables. One example is the global $21\,\textrm{cm}$ signal, which is a single spectrum of $21\,\textrm{cm}$ absorption or emission averaged over the entire angular area of the sky \citep{bowman2008toward, singh2018saras}. Recently, the Experiment to Detect the Global Epoch of reionization Step team (EDGES) reported a tentative detection of a $21\,\textrm{cm}$ absorption signature at $z \sim 17 $ \citep{bowman2018absorption}, although this result remains controversial \citep{2018Natur.564E..32H, 2018Natur.564E..35B, 2019ApJ...874..153B,2019ApJ...880...26S, 2020MNRAS.492...22S}. Global signal measurements are complemented by experimental efforts to map spatial fluctuations in the $21\,\textrm{cm}$ brightness temperature field. Most such efforts currently focus on a measurement of the power spectrum, i.e., the variance in Fourier space. Power spectrum measurements have the potential to significantly improve constraints on cosmological and astrophysical parameters of reionization models, and to potentially even discover new fundamental physics (e.g., \citealt{mcquinn2006cosmological, pober2014next, greig201521cmmc,pober201564, kern2017emulating, greig2017simultaneously, hassan2017epoch, park2019inferring, ghara2020constraining}). Typically, these measurements are pursued by low-frequency radio interferometer arrays, such as the Murchison Widefield Array\footnote{\url{http://www.mwatelescope.org}} (MWA; \citealt{tingay2013murchison, bowman2013science}), the Low Frequency Array\footnote{\url{http://www.lofar.org}} (LOFAR; \citealt{van2013lofar}), the Donald C. Backer Precision Array for Probing the Epoch of Reionization\footnote{\url{http://eor.berkeley.edu}} (PAPER; \citealt{parsons2010precision}), the Hydrogen Epoch of Reionization Array\footnote{\url{https://reionization.org}} (HERA; \citealt{deboer2017hydrogen}), and the Square Kilometre Array\footnote{\url{https://www.skatelescope.org}} (SKA; \citealt{mellema2013reionization, koopmans2015advancing}). Although no experiment has yet to claim a detection of the $21\,\textrm{cm}$ power spectrum at redshifts relevant to the EoR, steady progress has been made in recent years in the form of increasingly stringent and robust upper limits\citep{dillon2014overcoming, dillon2015empirical, beardsley2016first, patil2017upper, barry2019improving, kolopanis2019simplified, li2019first, mertens2020improved, trott2020deep}.

In this paper, we tackle the crucial problem of error estimation in the context of $21\,\textrm{cm}$ power spectrum measurements. While an extensive literature on power spectrum error estimation exists for CMB measurements and galaxy surveys, there are several challenges that are unique to $21\,\textrm{cm}$ cosmology. Chief amongst these is the fact that any measured signals will be strongly contaminated by the foregrounds, which are generally $4$ to $5$ orders of magnitude stronger in temperature \citep{de2008model,jelic2008foreground,bernardi2009foregrounds}. To overcome this obstacle, some collaborations pursue a strategy of foreground subtraction, where models of foreground emission are subtracted from the data (e.g., \citealt{harker2009non, bernardi2011subtraction, cho2012technique, chapman2012foreground, shaw2015coaxing}). Different approaches to foreground subtraction make different assumptions (see \citealt{liu2019data} for examples), but all face the same problem of attempting to subtract a large contaminant from a large raw signal to reveal a small cosmological signature. With empirical constraints on the low-frequency radio sky being relatively scarce and generally imprecise, the chances of mis-subtraction are high. Errors in such a subtraction process as well as the effects of subtraction residuals must therefore be propagated through to a final power spectrum estimate.

In this paper, however, we do not tackle the problem of error propagation in the context of foreground subtraction; instead, we consider error estimation in the context of foreground avoidance, where one aims to make cosmological measurements exclusively in Fourier modes where foregrounds are expected to be subdominant. Key to this is the notion of the foreground wedge, a regime in Fourier space beyond which spectrally smooth foregrounds cannot extend if observed using an ideal interferometer \citep{datta2010bright,parsons2012per,vedantham2012imaging,morales2012four,trott2012impact,thyagarajan2013study,hazelton2013fundamental,liu2014epoch}. The limitation of foregrounds to the wedge is a theoretically robust notion \citep{liu2019data}, and in principle one can make foreground-free measurements simply by avoiding the regime. In practice, observations are never made using perfect interferometers, and instrumental systematics such as having non-identical antenna elements, cable reflections, and cross couplings (e.g., \citealt{kern2019mitigating, kern2020mitigating}) complicate one's foreground mitigation efforts. These complications can result in the appearance of contaminants outside of the foreground wedge, and in this paper we define and tackle the problem of error estimation in two regimes: a noise-dominated regime and a signal-dominated regime (whether these signals could be foregrounds, systematics, or any other coherent signals).

Through a combination of analytic work, simulations of toy models, and tests on small amounts of real data, we critically examine different ways in which one can place error bars on 21\,cm delay power spectra. Our goal is to produce a ``buyer's guide" that enumerates the advantages and disadvantages of various error estimation methods. Understanding these strengths and weaknesses are crucial for setting upper limits, diagnosing systematics, interpreting the results of null tests, and for the design and optimization of future telescopes \citep{morales2005power,mcquinn2006cosmological, parsons2012sensitivity}. Although we will focus primarily on the delay power spectrum-style analysis \citep{parsons2012per} in support of recent HERA upper limits \citep{hera2021}, we expect many of our results to be more generally applicable.

This paper is organized as follows: in Section~\ref{sec:ps}, we review the basics of power spectrum estimation using the delay spectrum technique, establishing our notation. In Section~\ref{sec:error} we propose several methods for estimating errors in $21\,\textrm{cm}$ delay power spectra. These approaches are then compared and contrasted using simulations and real data in Section~\ref{sec:tests}. We then discuss the strengths and weaknesses of each error estimation method in Section~\ref{sec:discussion} before summarizing our conclusions in Section~\ref{ref:conc}. For readers' convenience, we provide dictionaries for a number of quantities defined in this paper in Tables \ref{tab:sca+fun} and \ref{tab:vec+mat}.

\section{Power Spectrum Estimation via the Delay Spectrum}
\label{sec:ps}
%%%
In this section we review the delay spectrum approach to $21\,\textrm{cm}$ power spectrum estimation \citep{parsons2012per} using the the language of the quadratic estimator (QE) formalism \citep{liu2011method} that we adopt in this paper. 

The delay spectrum technique enables power spectra to be estimated using just a single baseline of a radio interferometer, with fluctuations in the $21\,\textrm{cm}$ signal probed primarily in the line-of-sight direction via spectral information. The starting point is the visibility $V (\bm{b},\nu)$ measured by an interferometer's baseline $\bm{b}$ at frequency $\nu$. Under the flat-sky limit, it is given by
\begin{equation}
\label{eq:visibility}
    V(\bm{b},\nu) = \int I(\bm{\theta},\nu) A(\bm{\theta},\nu)\exp{\left(-i 2\pi \frac{\nu}{c}\bm{b}\cdot\bm{\theta}\right)} \text{d}^2\theta\,,
\end{equation}
where $c$ is the speed of light, $\bm{\theta}$ is the angular sky position, $I(\bm{\theta},\nu)$ is the source intensity function, and $A(\bm{\theta},\nu)$ is the primary beam function. If we express $I(\bm{\theta},\nu)$ in terms of its Fourier transform $\tilde{I}(\bm{u},\eta)$, i.e., 
\begin{equation}
\label{eq:intensity}
    I(\bm{\theta},\nu) = \int \tilde{I}(\bm{u},\eta) e^{i2\pi(\bm{u}\cdot\bm{\theta}+\eta\nu)} \text{d}^2u\text{d}\eta,
\end{equation}
then our visibility equation becomes
\begin{eqnarray}
\label{eq:visbility2Fourier}
    V(\bm{b},\nu) & = &  \int \tilde{I}(\bm{u},\eta)   A(\bm{\theta},\nu) e^{i2\pi(\bm{u}\cdot\bm{\theta}+\eta\nu-\bm{b}_\lambda\cdot\bm{\theta})} \text{d}^2u\text{d}\eta \text{d}^2\theta \notag \\
    & = &  \int \tilde{I}(\bm{u},\eta)   \tilde{A}(\bm{b}_\lambda-\bm{u} ,\nu) e^{i2\pi\eta\nu} \text{d}^2u\text{d}\eta,
\end{eqnarray}
where we have defined $\bm{b}_\lambda \equiv \frac{\nu}{c}\bm{b}$ as the normalized baseline vector for baseline $\bm{b}$ in units of wavelength. In the angular directions, we see that a visibility has a response to $\bm{u}$ modes centred around $\bm{b}_\lambda$. If the primary beam $A$ is fairly broad, $\tilde{A}$ will be highly compact and the majority of the integral will be sourced from $\bm{u} \approx \bm{b}_\lambda$. We will use this fact later. 
From this, one sees that a visibility $V(\bm{b}, \nu)$ is a linear function of $\tilde{I}(\bm{u},\eta)$. This quantity is directly related to the cylindrical power spectrum $P(\bm{u}, \eta)$, which decomposes power into Fourier wavenumbers perpendicular to the line of sight ($\bm{u}$) and parallel to the line of sight ($\eta$), and is formally defined as
\begin{equation}
\label{eq:power_spectrum_def}
    \langle\tilde{I}^*(\bm{u},\eta) \tilde{I}(\bm{u}',\eta') \rangle \equiv \delta^\text{D}(\bm{u}-\bm{u}')\delta^\text{D}(\eta - \eta') P(\bm{u},\eta).
\end{equation}
Such a power spectrum can be recast into more conventional cosmological coordinates via the relations\footnote{In addition to mapping the arguments of $P$, there is also an additional multiplicative constant; see \citet{liu2014epoch} for explicit expressions.}
\begin{equation}
\label{eq:u_eta2k}
\bm{k_\perp} = \frac{2\pi \bm{u}}{D_\text{c}}\,;\quad k_\parallel = \frac{2\pi\nu_{21}H_0 E(z)}{c(1+z)^2}\eta,
\end{equation}
where $D_\text{c}$ is the line-of-sight comoving distance, $\nu_{21}$ is the rest frequency of the $21\,\textrm{cm}$ line, $H_0$ is the Hubble parameter today, and $E(z) \equiv \sqrt{\Omega_\Lambda + \Omega_m (1+z)^3}$, with $\Omega_\Lambda$ and $\Omega_m$ as the normalized dark energy and matter density, respectively.

Since the power spectrum is a quadratic function of the Fourier representation of the sky, we expect that one should be able to estimate the power spectrum by forming some quadratic function of visibilities.
However, directly squaring some functions of the visibilities will incur a noise bias because noise that is symmetrically distributed about zero will have a positive contribution that does not average down with cumulative samples. Fortunately, the noise bias can be avoided by cross-multiplying nominally identical measurements rather than by squaring a single measurement. For instance, one might choose to form quadratic combinations of data from adjacent time samples of a single baseline's time stream, or perhaps to cross-multiply the time streams from two redundant baselines that satisfy $\bm{b}_1 = \bm{b}_2 = \bm{b}$ for some $\bm{b}$. In this paper, we will consider power spectrum measurements that are formed from cross-multiplications in \emph{both} time and different copies of an identical baseline. Utilizing both types of cross-multiplications has the advantage of avoiding skewness in the probability distributions of the measured power spectra, simplifying the interpretation of our results. This is discussed in Appendix~\ref{sec:crosscorrskewness}. In this section, however, we will---for simplicity---suppress explicit reference to the data time stream and use notation that explicitly refers to cross-correlating different baselines. Given a pair of redundant baselines  $\bm{b}_1$ and $\bm{b}_2$, we stack their measuring visibilities at multiple frequencies $\nu_1, \nu_2 ...$ at single time instants into two data vectors $\bm{x}_1$ and $\bm{x}_2$, such that 
\begin{equation}
\label{eq:data_vector}
\bm{x}_1 = \left( 
\begin{array}{c}
     V(\bm{b}_1,\nu_1) \\
     V(\bm{b}_1,\nu_2) \\
     \vdots
\end{array}\right); \quad
\bm{x}_2 = \left( 
\begin{array}{c}
     V(\bm{b}_2,\nu_1) \\
     V(\bm{b}_2,\nu_2) \\
     \vdots
\end{array}\right)\,.
\end{equation}

To make an explicit connection between visibilities and power spectra, we must examine the statistical properties of these data vectors. For quadratic statistics the key quantity is the covariance matrix $\bm{C}^{12}\equiv \langle\bm{x}_1 \bm{x}_2^\dagger\rangle$, which can be written as
\begin{eqnarray}
\label{eq:covariance_decompose_step0}
    \bm{C}^{12}_{ij} & \equiv &  \langle V(\bm{b_1},\nu_i) V^*(\bm{b_2},\nu_j)\rangle \notag \\
    & = &  \int P(\bm{u},\eta) \tilde{A}(\bm{b}_{\lambda1i}-\bm{u} ,\nu_i)\tilde{A}^*(\bm{b}_{\lambda2j}-\bm{u} ,\nu_j) \notag \\
    &\phantom{=} & \times e^{i2\pi\eta(\nu_i-\nu_j)} \text{d}^2u\text{d}\eta \notag \\
    & \approx &  \int P(\overline{\bm{b}}_\lambda,\eta) e^{i2\pi\eta(\nu_i-\nu_j)}\text{d}\eta \notag \\
    & \phantom{=} & \times \int \tilde{A}^*(\bm{b}_{\lambda1i}-\bm{u} ,\nu_i)\tilde{A}(\bm{b}_{\lambda2j}-\bm{u} ,\nu_j)\text{d}^2u \,, 
\end{eqnarray}
where $\bm{b}_{\lambda1i}$ and $\bm{b}_{\lambda2j}$ are the normalized baseline vectors for baseline $\bm{b}_1$ and $\bm{b}_2$ evaluated at frequencies $\nu_i$ and $\nu_j$, respectively, and $\overline{\bm{b}}_\lambda$ is the mean of the two. In deriving Equation \eqref{eq:covariance_decompose_step0},
we first substituted Equation \eqref{eq:visbility2Fourier} for the expressions of visibilities in the angle bracket, and then factored the evaluated cylindrical power spectrum out of the integral over $\bm{u}$. Next we replace the continuous integral on power spectra with discrete sums over a series of piecewise constant bandpowers $P(\overline{\bm{b}}_\lambda,\eta_\alpha)$, such that
\begin{eqnarray}
\label{eq:covariance_decompose_step1}
   \bm{C}^{12}_{ij}& \approx & \sum_\alpha P(\overline{\bm{b}}_\lambda,\eta_\alpha)  \int_{\eta_\alpha} e^{i2\pi\eta_\alpha(\nu_j-\nu_i)} \text{d}\eta \notag \\
    & \phantom{=} &\times \int \tilde{A}(\bm{b}_{\lambda1i}-\bm{u} ,\nu_i)\tilde{A}^*(\bm{b}_{\lambda2j}-\bm{u} ,\nu_j)\text{d}^2u \notag \\
    & \approx & \sum_\alpha P(\overline{\bm{b}}_\lambda,\eta_\alpha) e^{i2\pi\eta_\alpha(\nu_i-\nu_j)}\Delta \eta \notag \\
    & \phantom{=}& \times \int e^{-i 2 \pi (\bm{b}_{\lambda1i} - \bm{b}_{\lambda2j}) \cdot \bm{\theta}} A(\theta,\nu_i) A^*(\theta ,\nu_j)\text{d}^2 \theta \notag \\
    & \equiv & \sum_\alpha P(\overline{\bm{b}}_\lambda,\eta_\alpha) \bm{Q}^{12,\alpha}_{ij},
\end{eqnarray}
Henceforth, we will adopt the notation $P_\alpha \equiv P(\overline{\bm{b}}_\lambda,\eta_\alpha)$ to mean the value of the cylindrical power spectrum $P(\bm{u},\eta)$ evaluated at $\bm{u} = \overline{\bm{b}}_\lambda$ and $\eta = \eta_\alpha$. The index $\alpha$ discretely runs over a series of bins in $\eta$, and as long as these bins are narrow compared to the scales over which the power spectrum changes, a piecewise constant treatment is appropriate.

\begin{table*}[!htbp]
    \centering
    \begin{tabular}{c p{12cm} c}
    \toprule
         Quantity &  Definition/Meaning  & First Appearance\\
    \midrule
        $\bm{b};\,\bm{b}_p$ & Baseline vector; Vector of the $p$th index baseline & Equation \eqref{eq:visibility} \\
        $\bm{\theta}$ & Angular sky position & Equation \eqref{eq:visibility} \\
        $\nu$;\,$\nu_i$ & Frequency; Frequency of the $i$th index channel & Equation \eqref{eq:visibility} \\
         $\bm{b}_\lambda;\, \bm{b}_{\lambda pi}$ & Normalized baseline vector in units of wavelength; Normalized vector for baseline $\bm{b}_p$ at frequency $\nu_i$ & Equation \eqref{eq:visbility2Fourier}  \\
        $\bm{u}$ & Fourier dual to $\bm{\theta}$ & Equation \eqref{eq:intensity} \\  
        $\eta;\,\eta_\alpha$ & Fourier dual to $\nu$; the $\alpha$th index $\eta$ mode & Equation \eqref{eq:intensity} \\
        $\tau;\,\tau_\alpha$ & Delay, i.e., Fourier dual to $\nu$ on a single baseline; the $\alpha$th index delay mode & Equation \eqref{delay_spectra_decompose} \\ 
    \bottomrule
        $A(\bm{\theta}, \nu)$ & Primary beam function at position $\theta$ and frequency $\nu$ & Equation \eqref{eq:visibility} \\
        $\tilde{A}(\bm{u}, \nu)$ & Spatial Fourier Transform Dual of primary beam function & Equation \eqref{eq:visbility2Fourier} \\ 
        $\gamma(\nu)$ & Spectral tapering function at frequency $\nu$& Equation
        \eqref{eq:delay_transform}\\
        $N_\text{time};N_\text{blp}$ & Number of time instants; Number of baseline-pairs & Equation \eqref{eq:pavg} \\
         $N_\text{boot}$ & Number of bootstrapping sample sets & Equation \eqref{eq:bootstrap} \\
        $I(\bm{\theta}, \nu)$ & Sky source intensity function at position $\theta$ and frequency $\nu$ & Equation \eqref{eq:visibility} \\
        $\tilde{I}(\bm{u}, \eta)$ & Fourier transform of $I$ at angular wavenumber $\bm{u}$ and line-of-sight wavenumber $\eta$ & Equation \eqref{eq:intensity} \\
        $V(\bm{b}, \nu)$ & Visibility measured by baseline $\bm{b}$ at frequency $\nu$ & Equation \eqref{eq:visibility} \\
        $P(\bm{u}, \eta)$ & Cylindrical power spectrum at angular wavenumber $\bm{u}$ and line-of-sight wavenumber $\eta$ & Equation \eqref{eq:power_spectrum_def} \\
        $P_\alpha$ & The $\alpha$th bandpower $P_\alpha \equiv P(\overline{\bm{b}}_\lambda,\eta_\alpha)$ & Equation \eqref{eq:covariance_decompose_step1} \\
        $\hat{P}_\alpha$ & The estimator for the $\alpha$th bandpower $P_\alpha$& Equation \eqref{eq:palpha_estimator} \\
        $M_\alpha$ & The normalization scalar of the estimator for the $\alpha$th bandpower & Equation \eqref{eq:E_choice} \\
        $\tilde{V}(\bm{b}_p,\tau_\alpha), \tilde{x}_p(\tau_\alpha)$ & Delay spectra of baseline $\bm{b}_p$ at delay mode $\tau_\alpha$ & Equation \eqref{eq:delay_spectra_product}\\
        $\tilde{V}_\text{signal} (\bm{b}_p, \tau_\alpha), \tilde{s}_p(\tau_\alpha)$ & The signal component of $\tilde{V}$ of baseline $\bm{b}_p$ at delay mode $\tau_\alpha$ & Equation \eqref{delay_spectra_decompose}\\
         $\tilde{V}_\text{noise} (\bm{b}_p, \tau_\alpha), \tilde{n}_p(\tau_\alpha)$ & The noise component of $\tilde{V}$ of baseline $\bm{b}_p$ at delay mode $\tau_\alpha$& Equation \eqref{delay_spectra_decompose}\\
         $P_{\tilde{x}_1\tilde{x}_2}$ & Power spectra formed from visbilities $\bm{x}_1$ and $\bm{x}_2$ & Equation \eqref{eq:P_SN} \\
        \bottomrule
    \end{tabular}
    \caption{Dictionary of highlighted scalars and functions.}
    \label{tab:sca+fun}
\end{table*} 

Equation \eqref{eq:covariance_decompose_step1} shows the cross-baseline covariance matrix of visibilities encodes information about the power spectrum bandpowers via a family of response matrices $\bm{Q}^{12,\alpha}$ (with a different matrix for every value of the bandpower index $\alpha$). Since the covariance is an ensemble-averaged quadratic function of the data, one might venture that estimators for the bandpowers can be constructed by forming quadratic combinations of the data, i.e.,
\begin{equation}
\label{eq:palpha_estimator}
    \hat{P}_\alpha = \bm{x}_1^\dagger \bm{E}^{12,\alpha}\bm{x}_2\,,
\end{equation}
where $\bm{E}^{12,\alpha}$ is a matrix that can be chosen (within certain limitations) by the data analyst. Taking the ensemble average on both sides and inserting Equation \eqref{eq:covariance_decompose_step1} then yields
\begin{equation}
\label{eq:window_function_def}
\langle \hat{P}_\alpha \rangle = \sum_\beta \textrm{tr}\left( \bm{E}^{12,\alpha} \bm{Q}^{21,\beta} \right) P_\beta \equiv \sum_\beta W_{\alpha \beta} P_\beta,
\end{equation}
where $\bm{W}$ is the window function matrix. To ensure that our estimated bandpowers are correctly normalized, we require that each row of $\bm{W}$ sum to unity. 

In the HERA power spectrum pipeline, we pick a family of $\bm{E}^{12}$ matrices of the form
\begin{equation}
\label{eq:E_choice}
\bm{E}^{12,\alpha} \equiv M_{\alpha} \bm{R}_1 \bm{Q}^{\text{DFT}, \alpha} \bm{R}_2,
\end{equation}
where the matrix $\bm{Q}^{\text{DFT}, \alpha}_{ij} \equiv  e^{i 2 \pi \eta_\alpha (\nu_i - \nu_j)}$ is responsible for taking the Fourier transform of the two copies of the data vectors in the quadratic estimator. The matrices $\bm{R}_1$ and $\bm{R}_2$ are weighting matrices that act on visibilities from $\bm{b}_1$ and $\bm{b}_2$, respectively. In this paper, we use $\bm{R}= \bm{T}\bm{Y}$, where both $\bm{T}$ and $\bm{Y}$ are diagonal matrices. The former is used to impose a Blackman-Harris tapering function on the spectral data, and the latter propagates data flags. With a quadratic estimator of this form, the normalization scalar, $M_\alpha$, should take the form    
\begin{equation}
\label{eq:M}
M_\alpha = \frac{1}{ \sum_{\beta} \text{tr}(\bm{R}_1 \bm{Q}^{\text{DFT}, \alpha} \bm{R}_2 \bm{Q}^{12,\beta})}\,
\end{equation}
which ensures that the rows of $\mathbf{W}$ sum to unity, and therefore that the bandpowers are properly normalized. In our case, we do use this normalization, but we approximate the $\bm{Q}^{12,\beta}$ term in the denominator. Rather than evaluating the full integral in Equation \eqref{eq:covariance_decompose_step1}, we make the approximation that $\mathbf{b}_{\lambda1i} \approx \mathbf{b}_{\lambda2i}$. In fact, this is the motivation for the use of $\bm{Q}^{\text{DFT},\alpha}$ in Equation~\eqref{eq:E_choice} rather than $\bm{Q}^{12}$; notice that if $\mathbf{b}_{\lambda1i} = \mathbf{b}_{\lambda2i}$, then $\bm{Q}^{12} \propto \bm{Q}^{\text{DFT}}$. Over large bandwidths, this will fail for long baselines, since $\bm{b}_\lambda \equiv \nu\bm{b} / c$. 

The approximation that we have just made is equivalent to the delay spectrum approximation \citep{parsons2012per, liu2014epoch}. To see this, we can write our estimator in the continuous limit. Our current form for $\bm{E}^{12,\alpha}$ is separable into the product of two matrices that each involve only one of the two baselines. In particular, if $\gamma(\nu)$ is the functional form of the Blackman-Harris taper, then we have $\bm{E}^{12,\alpha}_{ij} = \gamma_1(\nu_i) e^{i 2 \pi \eta_\alpha (\nu_i-\nu_j) } \gamma_2(\nu_j)$, and its action on each baseline's visibilities in Equation \eqref{eq:palpha_estimator} is to compute the quantity
\begin{equation}
\sum_{i} V(\bm{b},\nu_i) \gamma(\nu_i) e^{-2\pi \eta \nu_i} \Delta \nu,
\end{equation}
which is just a discrete approximation to
\begin{equation}
\label{eq:delay_transform}
\tilde{V}(\bm{b},\eta)  = \int V(\bm{b},\nu) \gamma(\nu) e^{-i 2\pi \eta \nu} \text{d}\nu \,.
\end{equation}
Note Equation \eqref{eq:delay_transform} is an equivalent expression of the delay transform in \citet{parsons2012per}. Therefore
\begin{eqnarray}
\label{eq:delay_spectra_product}
    \hat{P}_\alpha & = & \bm{x}_1^\dagger \bm{E}^{12,\alpha}\bm{x}_2 \notag \\ & \propto & \sum_{ij} V^*(\bm{b}_1,\nu_i) \gamma_1(\nu_i) V(\bm{b}_2,\nu_j) \gamma_2(\nu_j)e^{i 2 \pi \eta_\alpha (\nu_i - \nu_j)} \notag \\
    & = & \tilde{V}^*(\bm{b}_1,\eta_\alpha) \tilde{V}(\bm{b}_2,\eta_\alpha) \,.
\end{eqnarray}
Equation \eqref{eq:delay_spectra_product} just indicates that the quadratic estimator is proportional to the product of delay-transformed visibilities. This is an estimator that is based on Fourier transforming the visibility spectra from individual baselines, rather than combining information from different baselines. In principle, only the latter can probe truly rectilinear Fourier modes on the sky, since $\bm{k}_\perp \propto \bm{b}_\lambda$ (which is a frequency-dependent quantity), and thus to probe the same $\bm{k}_\perp$ at multiple frequencies---which is needed to perform the Fourier transform along the line-of-sight direction---one needs multiple baselines. The delay spectrum approach uses the fact that $\bm{b}_\lambda$ evolves only slowly with frequency for short baselines to form an approximate power spectrum estimator. We make this approximation throughout this paper, as this is the choice that has been made for the next iteration of power spectrum upper limits from HERA observations. In recognition of this, we will henceforth use $\tau$ to index our line-of-sight Fourier modes (as is customary for delay spectra) instead of $\eta$ (which is generally used to denote true rectilinear line-of-sight wavenumbers) \citep{morales2012four,2019MNRAS.483.2207M}.

\begin{table*}[!htbp]
    \centering
    \begin{tabular}{c p{12cm} c c}
    \toprule
    Quantity &  Definition/Meaning  & Size & First Appearance \\
    \midrule
       $\bm{x}_p$ & Stacked visibilities at multiple frequencies of baseline $\bm{b}_p$ & $N_\text{freq}$ & Equation \eqref{eq:data_vector} \\
    \bottomrule
        $\bm{C}^{pq}$ & Covariance matrices $\bm{C}^{pq}\equiv \langle\bm{x}_p \bm{x}_q^\dagger\rangle$ & $N_\text{freq} \times N_\text{freq}$ & Equation \eqref{eq:covariance_decompose_step0}  \\
        $\bm{Q}^{pq,\alpha}$ & Response of covariance  $\bm{C}^{pq}$ to the $\alpha$th bandpower & $N_\text{freq} \times N_\text{freq}$ & Equation \eqref{eq:covariance_decompose_step1}\\
        $\bm{E}^{pq,\alpha}$ & Matrix for quadratic estimator of bandpower $P_\alpha$, i.e., $\hat{P}_\alpha = \bm{x}_p^\dagger \bm{E}^{pq,\alpha}\bm{x}_q$ & $N_\text{freq} \times N_\text{freq}$ & Equation \eqref{eq:palpha_estimator} \\
        $\bm{W}$ & Window function matrix & $N_\text{delay} \times N_\text{delay}$ & Equation \eqref{eq:window_function_def} \\
        $\bm{R}_p$ & Weighting matrix acting on $\bm{x}_p$ & $N_\text{freq} \times N_\text{freq}$ & Equation \eqref{eq:E_choice} \\
        $\bm{Q}^{\text{DFT}, \alpha}$ & Matrix taking Fourier Transform in the estimator & $N_\text{freq} \times N_\text{freq}$ &
        Equation \eqref{eq:E_choice} \\
        $\bm{U}^{pq}$ & two-point correlation matrices $\bm{U}^{pq}\equiv \langle\bm{x}_p \bm{x}_q^T\rangle$ & $N_\text{freq} \times N_\text{freq}$ & Equation \eqref{eq:covariance} \\
         $\bm{G}^{pq}$ & two-point correlation matrices $\bm{G}^{pq}\equiv \langle \bm{x}_p^* \bm{x}_q^\dagger\rangle$ & $N_\text{freq} \times N_\text{freq}$ & Equation \eqref{eq:covariance} \\
    \bottomrule
    \end{tabular}
    \caption{Dictionary of highlighted vectors and matrices.}
    \label{tab:vec+mat}
\end{table*}

In the language of the delay spectrum, the foreground wedge becomes particularly simple to describe: smooth spectrum foregrounds simply contaminate all modes below a particular delay, the value of which depends on the baseline length \citep{parsons2012per,liu2014epoch,liu2019data}. Suppose we decompose the delay transformed visibility into the signal component $\tilde{V}_\text{signal}$ (mainly foregrounds, and we are neglecting the much weaker EoR signal here) and the noise component $\tilde{V}_\text{noise}$, such that
\begin{eqnarray}
\label{delay_spectra_decompose}
\tilde{V}(\bm{b}_1,\tau_\alpha)  &\equiv & \tilde{x}_1(\tau_\alpha) \nonumber \\
&\equiv& \tilde{V}_\text{signal}(\bm{b}_1,\tau_\alpha) + \tilde{V}_\text{noise}(\bm{b}_1,\tau_\alpha) \nonumber \\
&\equiv &\tilde{s}_1(\tau_\alpha)+ \tilde{n}_1(\tau_\alpha).
\end{eqnarray}
Since we are working on redundant baselines, we will henceforth drop the subscript on $\tilde{s}$, as the two baselines used in Equation \eqref{eq:delay_spectra_product} should measure identical signals. Mathematically, then, the statement that the smooth spectrum foregrounds contaminate only low delay modes is given by
\begin{equation}
\label{eq:wedge}
 \hat{P}_\alpha \approx \left\{ \begin{array}{ll}
\tilde{s}^*\tilde{s} + \tilde{s}^*\tilde{n}_2 + \tilde{n}_1^*\tilde{s}\, &\quad \textrm{if $|\tau_\alpha| < \tau_0$}\\
\tilde{n}_1^*\tilde{n}_2\, &\quad \textrm{otherwise},\\
\end{array} \right.
\end{equation}
where $\tau_\alpha$ is the delay corresponding to the $\alpha$th bandpower, and $\tau_0$ is some critical delay value that separates parts of the power spectrum that are foreground-dominated from those that are not. In general, $\tau_0$ will depend on the properties of one's instrument as well as the extent to which the assumption of smooth foregrounds is good. At delays less than $\tau_0$, we have assumed that the foreground signal is so large that the noise-noise cross term can be neglected.

Throughout the rest of this paper, we will appeal to Equation \eqref{eq:wedge} for intuition when contemplating the behaviour of our power spectrum estimates at different delays. For now, we note two of its important properties. First, while the power spectrum of a signal $\tilde{s}^*\tilde{s}$ will be always real valued, the overall estimator $\hat{P}_\alpha$ is complex. It is possible to write down symmetrized estimators that give real power spectra. However, since the imaginary part is sourced by noise, it is a useful diagnostic quantity to examine. Second, even though the noise-noise terms may be negligible in the signal dominated regimes, there will still be a considerable uncertainty here that enters via the signal-noise cross terms.

Until now, we have focused on power spectra estimated from visibilities measured at single time instants. Given data from multiple times, we can average the power spectra estimated from individual measurements together. For a drift scan telescope, this averaging of power spectra from different time samples is tantamount to invoking statistical isotropy to justify the spherical averaging of power spectra over different wavevector $\mathbf{k}$ directions. In addition to averaging in time, if we have multiple pairs of baselines within the same redundant group of baselines, we may average over the power spectrum estimates from multiple baseline pairs. The simplest way to do this is to perform an unweighted average:
\begin{equation}
\label{eq:pavg}
  \overline{\hat{P}}_\alpha = \frac{1}{N_\text{time} N_\text{blp}} \sum_{\text{time},\text{blp}} \hat{P}_\alpha(\text{time},\text{blp})\,,
\end{equation}
where $N_\text{time}$ is the number of time integrations, $N_\text{blp}$ is the number of baseline pairs, $ \hat{P}_\alpha(\text{time},\textrm{blp})$ is the power spectrum estimate (given by previous equations in this section) at a time instant and a baseline pair (``blp"), and $ \overline{\hat{P}}_\alpha$ is the average of estimates. The type of averaging performed here may be termed an ``incoherent average", to distinguish it from a ``coherent average", where one averages over visibilities (or converts them into a single image) before squaring them in power spectrum estimation. The latter provides greater sensitivity---if calibration errors and other systematic effects can be brought under control \citep{2019MNRAS.483.2207M}. The former retains the ability to inspect the contributions from particular baseline pairs and time until right before the final result, making some systematics easier to diagnose. However, note that by employing a suitable fringe-rate filtering of the time-stream data, it is in principle possible to recover the lost sensitivity from a ``square-then-add" approach \citep{parsons2016fringe}. In this paper, we will focus on the error statistics of the incoherent average approach, as this is what is currently used in the HERA pipeline \citep{hera2021}. 

Before we move into the discussion on error estimation methods in the next section, it is worth noting that Equation \eqref{eq:pavg} is not the optimal way to obtain average power spectra with the least variance. Generally, given a set of estimates $\hat{\bm{P}}_\alpha$ for bandpower $P_\alpha$ with measurement errors $\bm{\sigma}$, such that 
\begin{align}
\hat{\bm{P}}_\alpha & = \bm{D}P_\alpha  + \bm{\epsilon} \,,
\end{align}
an linear estimator of $P_\alpha$ is written as 
\begin{align}
\label{eq:palpha_avg_estimator}
\overline{\hat{P}}_\alpha & = \bm{K} \hat{\bm{P}}_\alpha\,.
\end{align}
Here $\bm{D}$ is a column vector of 1s. We need to select $\bm{K}$ such that $\bm{KD} = \bm{I}$ in order to achieve an unbiased constraint that satisfies $\langle \overline{\hat{P}}_\alpha \rangle = P_\alpha$. For an arbitrary matrix $\bm{K}$, the error bar $\Sigma_\alpha \equiv \langle |\overline{\hat{P}}_\alpha - P_\alpha|^2\rangle = \bm{K \epsilon} \bm{K}^t$, where the error covariance matrix $\bm{\epsilon} \equiv \langle\bm{\sigma} \bm{\sigma}^t \rangle$. The superscript ``$t$" used here and along in this paper refers to the matrix transposition. Note that Equation~\eqref{eq:pavg} is just a special case where $\bm{K} = [\bm{D}^t \bm{D}]^{-1}\bm{D}^t$. 
When $\Sigma_\alpha$ is minimized (optimal), $\overline{\hat{P}}_\alpha$ and the corresponding $\Sigma_\alpha$ should take the form of \citep{tegmark1997make, dillon2014overcoming}
\begin{align}
\label{eq:opt_estimator} 
    \overline{\hat{P}}_\alpha & = [\bm{D}^t \bm{\epsilon}^{-1} \bm{D}]^{-1}\bm{D}^t \bm{\epsilon}^{-1} \hat{\bm{P}}_\alpha \\
    \Sigma_\alpha & = [\bm{D}^t \bm{\epsilon}^{-1} \bm{D}]^{-1} \,,
    \label{eq:opt_estimator_variace}
\end{align}
which amounts to an inverse covariance weighting of the data in averaging it down. Equation \eqref{eq:opt_estimator} brings us the ability to propagate the full covariance information over samples to obtain an least-variance average result. The diagonal elements of $\bm{\epsilon}$ are easily interpreted as the variance in each individual measurement, while the off-diagonal elements, reflected by the coherency between time samples and baseline-pair samples, are far more complicated. If estimating the covariance matrix $\bm{\epsilon}$ of the pre-averaged data is difficult, one may opt to weight the data using some other matrix $\bm{\Gamma}$ instead of $\bm{\epsilon}$ in Equation \eqref{eq:opt_estimator}. In this case, the final variance $\Sigma_\alpha$ ends up being
\begin{equation}
    \Sigma_\alpha = [\bm{D}^t \bm{\Gamma}^{-1} \bm{D}]^{-1} \bm{D}^t \bm{\Gamma}^{-1} \bm{\epsilon} \bm{\Gamma}^{-t} \bm{D} [\bm{D}^t \bm{\Gamma}^{-t} \bm{D}]^{-1} \,.
\end{equation}
In principle, one could model the off-diagonal elements of $\bm{\epsilon}$. This is particularly important in the cosmic-variance dominated regime where the sky signal---which is what sources a cosmic variance error---is slowly drifting through HERA's field of view over the course of the day, thus inducing strong correlations between different time samples. In this paper we do not consider the modelling of off-diagonal covariances in $\bm{\epsilon}$ (or between different $\alpha$ values in $\overline{\hat{P}}_\alpha$). We assume diagonal covariance matrices and set $\Gamma = \bm{I}$, i.e., we use Equation~\eqref{eq:pavg} when computing the  ``incoherently-averaged'' power spectra, and here we are acknowledging other possibilities only for completeness.

\section{Error Estimation Methodology}
\label{sec:error}
%%%
\begin{table*}[!htbp]
    \centering
    \begin{tabular}{c p{12cm} c}
    \toprule
         Name &  Description & Definition \\
    \midrule
       $\sigma_\text{bs}$ & Error bar of the average power spectra by bootstrapping over the collection of samples & Equation \eqref{eq:bootstrap} \\
       $P_\text{diff}$ & Power spectra from differenced visibility used as a form of error bar & Equation \eqref{eq:P_diff} \\
       $P_\text{N}$ & Analytic noise power spectrum & Equation \eqref{eq:P_N} \\
       $P_\text{SN}$ & Error bar based on $P_\text{N}$ but including the extra signal-noise cross term  & Equation \eqref{eq:P_SN} \\
       $\sigma_\text{QE-N}$ & Error bar from the output covariance in QE formalism including only noise-noise term & Equation \eqref{eq:QE-N} \\
       $\sigma_\text{QE-SN}$ & Error bar from the output covariance in QE formalism including noise-noise term and signal-noise term & Equation \eqref{eq:QE-SN} \\
       $\tilde{P}_\text{SN}$ & Same as $P_\text{SN}$ but with an adjustment for noise double-counting & Equation~\eqref{eq:P_SN_bias} \\
       $\tilde{\sigma}_\text{QE-SN}$ & Same as $\sigma_\text{QE-SN}$ but with an adjustment for noise double-counting& Equation~\eqref{eq:C_signal} \\
    \bottomrule
    \end{tabular}
    \caption{Dictionary of error bars.}
    \label{tab:intro}
\end{table*}
Placing robust error bars on power spectra is crucial to our data analysis, whether it is for setting upper limits, diagnosing experimental systematics, or eventually declaring a detection of the cosmological $21\,\textrm{cm}$ signal.  Generally, contributions to the error bars of observed power spectra come from three sources: the EoR signal, noise, and foregrounds \citep{thyagarajan2013study,trott2014comparison,dillon2014overcoming, dillon2015empirical, lanman2019fundamental}. Of course, this is all complicated by the response of one's instrument, and ultimately, one's ability to place reliable error bars rests on one's ability to understand the behaviour of each data source in the context of the instrument.

The intrinsic variance of the EoR signal, also known as ``cosmic variance'', is the ensemble covariance on all possible realizations of the 21-cm temperature field. If the field is Gaussian, then its cosmic variance is proportional to the square of the power spectrum amplitude over the number of independent modes. \citet{lanman2019fundamental}, for example, estimate the cosmic variance could go as high as $\sim$ 35\% of the EoR signal for HERA-like fields of view with eight hours of local sidereal time (LST) observations using only the shortest (14.6-m) baselines of HERA. This uncertainty due to cosmic variance is brought down to a few percent level for the spherically averaged power spectrum when using all types of baselines. Importantly, as reionization evolves, the 21-cm temperature field is expected to become highly non-Gaussian, and the excess contribution from the non-Gaussian component could lift the cosmic variance in Gaussian part staggeringly, which is significant and should be considered for future high-sensitivity measurements \citep{mondal2015statistics, 2017MNRAS.464.2992M, 2019MNRAS.487.4951S}. In this paper, however, we assume that at our current levels of precision the cosmic variance is sub-dominant to noise and foregrounds.  

For instrumental noise, we assume that the noise in the visibility from each baseline is independent and Gaussian-distributed.
This is what one might expect based on the statistics of correlator outputs in a radio interferometer, but is also an assumption that we will see borne out in our empirical data in Section \ref{sec:tests}. With these well-understood statistical properties, the noise-dominated delays (recall Equation \ref{eq:wedge}) are relatively easy to model, at least in principle.

The low-delay, foreground-dominated regimes are trickier to model. One key problem is that the statistics of foregrounds are not well-understood, particularly at the low frequencies relevant to us. There are different approaches that one can take to this roadblock. The first is where one attempts to make a measurement of the cosmological $21\,\textrm{cm}$ signal only, by proactively subtracting (or simultaneously fitting) a foreground model. To properly set error bars on such a power spectrum, it is necessary to propagate uncertainties (accounting for the possibility of mis-subtractions) in the foreground model to the final errors (or in the case of a simultaneous fitting, to allow the errors on the cosmological signal to be appropriately inflated as one marginalizes over foreground uncertainties). While conceptually straightforward, these steps are difficult to implement in practice without a deep understanding of foreground statistics.

Instead, in this paper we treat foregrounds as additive systematics on the total sky emission. Crucially, this means we only require empirical knowledge of the foregrounds themselves, and not their full probability distribution. We simply quantify the error bars on a measurement of total sky emission due to instrumental noise, rather than what the error bars on the cosmological signal due to foreground uncertainties and noise. Some understanding of foregrounds is still needed for setting our errors because of the signal-noise cross terms in Equation~\eqref{eq:wedge}. Implicit in this approach is a strategy of foreground avoidance in the hunt for a cosmological signal detection, where it is hoped that the separation between foreground-dominated and foreground-negligible regimes in Equation~\eqref{eq:wedge} is a clean one. It is important to note, however, that we seek to compute error bars that transition smoothly between the regimes and are valid even if the conceptual separation is not a clean one in practice.\footnote{We stress that our analysis does not cease to apply at a certain delay---it is simply the case that at high delays, there is less of a pressing need to construct detailed models for foreground subtraction, which to some extent mitigates the need to consider the complicated statistical properties of this subtraction. It is likely that our formalism can be generalized to encompass some foreground subtraction, but detailed work beyond the scope of this paper would be necessary. As an example, suppose one were to use information at $\tau =0$ and an instrument model to subtract off leakage from other low (but non-zero) delay modes. In such a scenario, one would need to account for the fact that the noise contributions between different delay modes are now coupled. This can in principle be accommodated with appropriate covariance matrix modeling, but we leave this to future work.}

In addition to foregrounds, one can treat instrumental systematics in the same way. In other words, interpreting systematics as additive ``signals", the signal-noise cross term in the variance of power spectra is sourced by not just foregrounds, but also other systematics such as cable reflections and cross couplings \citep{kern2019mitigating, kern2020mitigating}. We can apply some models to remove systematics from the signal, but the residuals due to mis-subtraction will still increase the total uncertainties via the signal-noise cross term. Note, however, that in this paper we do not develop a comprehensive model to account for all systematics, which is particularly difficult when unknown modeling errors are present in complicated effects (e.g. direction-dependent gains). We will instead argue that a procedure of using the measured visibility itself to model the foregrounds and systematics allows us to set robust upper bounds, provided certain safeguards are in place to avoid biases. We will leave more exquisite \emph{a priori} characterizations of foregrounds and systematics in the signal-noise cross terms for the future.

Finally, one might worry that the averaging of power spectra from multiple measurements together like Equation \eqref{eq:pavg} might complicate the statistics. Appendix \ref{ap:bootstrap} shows an example of this. There, we show that when averaging over redundant baseline-pairs, the variance of average power spectra in the foreground-dominated regime goes down roughly with $N_\text{blp}^{-1/2}$ and not $N_\text{blp}^{-1}$ because some baselines will appear in multiple baseline \emph{pairs}. In other words, in foreground-dominated (or systematics-dominated) regimes, one cannot assume that baseline pairs average together in an independent fashion. This has consequences for certain methods of error bar computation, such as the bootstrapping approach discussed in the next subsection, which will tend to underestimate error bars in these regimes. To avoid this, one might just use \emph{pairs} in which each baseline only appears once in all baseline \emph{pairs}, or to compute a correction factor on the final results. In contrast to the foreground/signal-dominated regime, in the noise-dominated regime one obtains correct final error bars by assuming that the baseline-pair samples are independent (even if they are not for the aforementioned reasons). In this paper, to avoid averaging power spectra over correlated samples, we will concentrate on the averaging of power spectra of a single baseline-pair over multiple time samples.  

We will have a more extensive discussion of the meaning of our error bars in Section~\ref{sec:discussion}. For concreteness, however, we will now propose several different methods for generating error bars based on the HERA power spectrum pipeline before performing quantitative comparisons in Section~\ref{sec:tests}. For the convenience of our readers, we provide Table \ref{tab:intro} as a quick preview.

\subsection{Bootstrap}
\label{subsec:boot}
Bootstrapping is a natural method for computing the error bars on the final averaged power spectrum with only minimal \emph{a priori} modeling assumptions. Within the $21\,\textrm{cm}$ cosmology literature, it has previously been used to set error bars on power spectrum upper limits (\citealt{parsons2014new, ali201564}; although see \citealt{cheng2018characterizing} for caveats on these limits). Bootstrapping is a process that goes hand in hand with the averaging step described in Equation~\eqref{eq:pavg}. Rather than performing a single average, we repeatedly form a new set of pre-averaged data by resampling the original set with replacement (i.e., allowing repeated entries). A new estimate of the final average, $\overline{\hat{P}}^{(k)}$, can be produced from the $k$th draw. The scatter in the realizations of the final averaged power spectrum is then quoted as an error bar $\sigma_\text{bs}$, such that
\begin{equation}
\label{eq:bootstrap}
    \sigma^2_\text{bs} = \frac{1}{N_\text{boot}}\sum_k \left[\overline{\hat{P}}^{(k)} - \frac{1}{N_\text{boot}}\sum_l \overline{\hat{P}}^{(l)}  \right]^2 \,,
\end{equation}
where $N_\text{boot}$ is the number of bootstrapping sample sets. In essence, one is using the data itself as an empirical estimate of the distribution from which the data is drawn \citep{efron1994introduction, press2007numerical}.

If the input data samples are independent and identically distributed, bootstrapping will give the same error bars as the true ones from ensemble average. However, this assumption is likely to be violated with our data. Consider the two axes that we have at our disposal. One possibility is to bootstrap over different time samples. Over short timescales, different time integrations have relatively uncorrelated noise realizations. However, as our drift scan telescope moves across different local sidereal time (LST) values, the sky brightness seen by the telescope changes, leading to slow changes in the noise level for a sky-noise dominated telescope. An alternative to bootstrapping over time is to bootstrap over different copies of an identical (``redundant") baseline group. Here, the downside is that it remains an open question as to how truly redundant current interferometric arrays are \citep{dillon2020redcal}, and precisely what the consequences of non-redundancy are \citep{Choudhuri2021}.

With correlated data samples, bootstrapping tends to underestimate the true error bars on a final averaged power spectrum \citep{cheng2018characterizing}. On the other hand, non-stationary effects such as non-redundancy can inflate bootstrap errors rather than revealing the fact that the data in fact come from multiple distributions. In later sections, we will compute error bars that come from bootstrapping over different LSTs, but will interpret these results with caution given the caveats we have just outlined. Of course, these caveats by no means diminish the value of bootstrap errors as yet another consistency check, particularly when one is diagnosing systematic effects (e.g., \citealt{kolopanis2019simplified}).

\subsection{Direct Noise Estimation By Visibility Differencing}
\label{subsec:P_diff}
\begin{figure*}[htbp!]
    \centering
    \includegraphics[width=\linewidth]{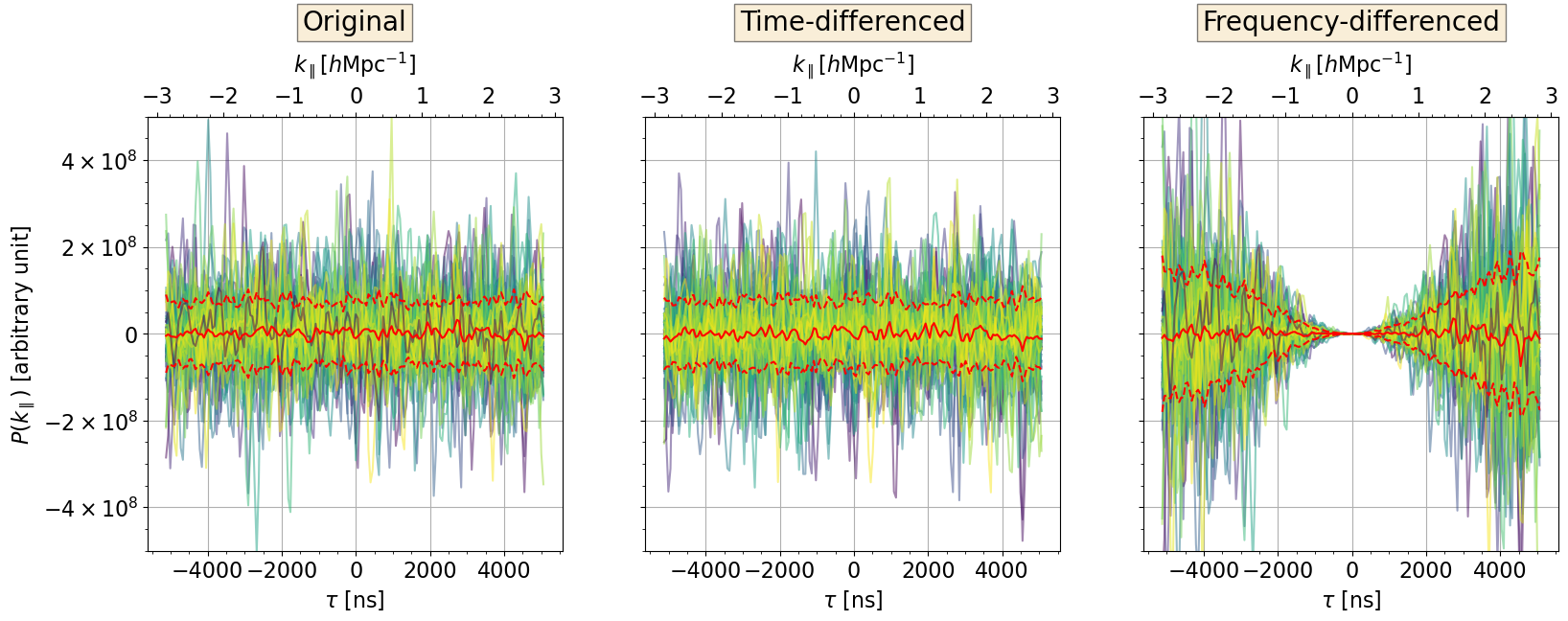}
    \caption{Here we generate $\sim 60$ realizations of time streams of white-Gaussian-noise visibilities, and compute the time-differenced visibilities and frequency-differenced visibilities respectively. \textit{Left}: Power spectra from the original visibilities.  \textit{Center}: Power spectra from time-differenced visibilities. \textit{Right}: Power spectra from frequency-differenced visibilities. In each panel, we plot the power spectra from every realization, along with the mean (solid red) and the standard deviation (dashed red) of power spectra over all realizations. We see power spectra from frequency-differenced visibilities are highly suppressed at low delays.}
    \label{fig:freq_difference}
\end{figure*}

The foreground and EoR signal varies relatively slowly in time (or frequency), such that after differencing the integrated visibility between very close LSTs (or frequencies), the normalized residual,
\begin{align}
\label{eq:visibility_diff}
    V_\text{diff} = & \frac{V(\bm{b},\nu,t_1) -  V(\bm{b},\nu,t_2)}{\sqrt{2}} \notag \\
    & \hspace{1.5cm} \text{or}  \notag \\
     V_\text{diff} = & \frac{V(\bm{b},\nu_1, t) -  V(\bm{b},\nu_2, t)}{\sqrt{2}} \,,
\end{align}
is almost noise-like. We can propagate such $ V_\text{diff}$ through power spectrum estimation pipelines to generate a ``noise-like" power spectrum $P_\text{diff}$, such that
\begin{equation}
\label{eq:P_diff} 
P_\text{diff} \propto \tilde{V}^*_\text{diff} \tilde{V}_\text{diff} \,,
\end{equation}
where appropriate proportionality/normalization constants allow $P_\text{diff}$ to have the same units as---and therefore be directly comparable to---power spectra. This quantity can be viewed as a random variable that represents random realizations of the noise in the system, which can be used to at least roughly estimate error bars in noise-dominated regimes (see Appendix \ref{ap:time_diff} for more details). It can be computed from either time-differenced or frequency-differenced visibilities. However, by differencing neighbouring points in frequency, we are in fact applying a high-pass filter in the delay space, which means that power is suppressed at low delay modes. This is illustrated in Figure \ref{fig:freq_difference}, and for this reason that the time-differencing method is preferred for empirical noise uncertainty estimation. However, it is important to note that many correlators do not dump data to disk fast enough for this to be feasible, as the sky changes non-negligibly on the timescale of a few seconds. The maximum time length of a single integration before reaching a decorrelation threshold depends on the baseline length, thus ones need particular simulations for their instruments to determine the suitable time scale \citep{2018MNRAS.476.2029W}. For the upgraded HERA correlator, it will be able to produce time-differenced visibilities on the milli-second timescale for accurate, empirical noise estimates. 

\subsection{Power Spectrum Method}
\label{subsec:ps_method}
With appropriate approximations (see \citealt{liu2019data} for details), it is possible to write down an analytic expression for the noise power spectrum given a system temperature, $T_\text{sys}$ in units of Kelvin:
\begin{equation}
\label{eq:P_N}
    P_\text{N} = \frac{X^2 Y \Omega_\text{eff} T_\text{sys}^2}{t_\text{int} N_\text{coherent}\sqrt{2N_\text{incoherent}}}\,,
\end{equation}
where $X\equiv D_\text{c}$ and $Y \equiv \frac{c (1+z)^2}{\nu_{21} H_0 E(z)}$ are conversion factors from sky angles and frequencies to cosmological coordinates, $\Omega_\text{eff}$ is the effective beam area, $t_\text{int}$ is the integration time, $N_\text{coherent}$ is the number of samples averaged at the level of visibility while $N_\text{incoherent}$ is the numbers of samples averaged at the level of power spectrum \citep{ZFH2004,pober2013, cheng2018characterizing, kern2020mitigating}.
This is an estimate of the root-mean-square (RMS) of a power spectrum measurement in the limit that it is purely thermal noise dominated. The system temperature, $T_\text{sys} = T_\text{sky} + T_\text{rcvr}$, is the sum of the sky and receiver temperature and describes the total noise content of the visibilities formed between cross-correlating data from different antennas \citep{Thompson2017}.

There are many ways in which the key quantity $T_{\rm sys}$ can be estimated. For example, we can take advantage of the differenced visibilities discussed in the previous subsection. These differences can then be converted into an estimate of $T_{\rm sys}$ via the relation
\begin{equation}
\label{eq:RMS_Tsys}
V_\text{RMS}(\{p,q\}) = \frac{2k_b \nu^2\Omega_p}{c^2}\frac{T_\text{sys,\!\{p,q\}}}{\sqrt{B \Delta t}} \,,
\end{equation}
where $k_b$ is the Boltzmann constant, $\Omega_p$ is the integrated beam area, $B$ is the bandwidth, and $\Delta t$ is the integration time at a single time sample. The ``RMS" subscript signifies taking the root-mean-square of the differenced visibilities and $p$ and $q$ are indices denoting two different antennas that form a baseline $\{p,q\}$. This serves to emphasize the fact that we can have a distinct system temperature for every baseline.

\begin{figure*}[htbp!]
    \centering
    \includegraphics[width=0.6\linewidth]{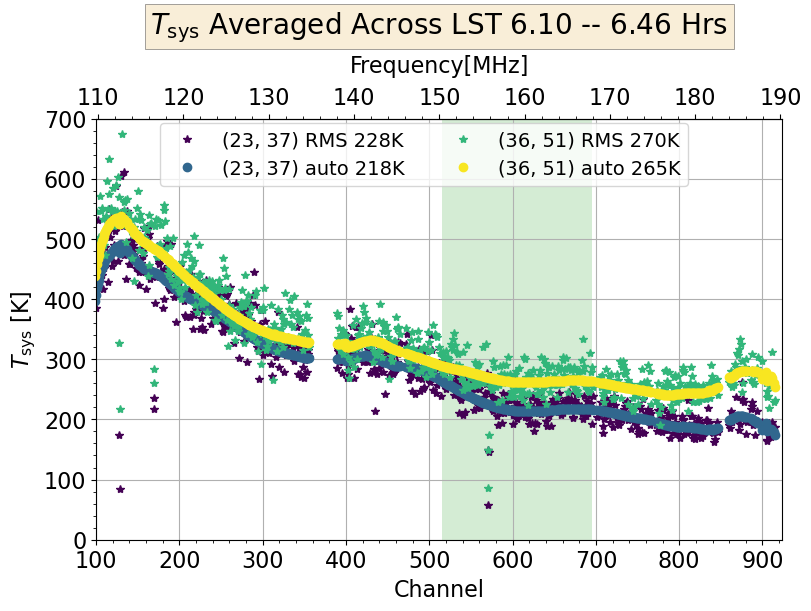}
    \caption{Comparison of two ways to estimate the system temperature based on HERA data. The system temperatures of cross-correlation visibilities on two $14.6\,\textrm{m}$ baselines [indexed by HERA antenna numbers (23, 37) and (36, 51)] are averaged across the LST range of 6.10 to 6.46 hours. The green regime, from frequency channel number \#515 to 695, show the HERA data band used for analysis in this paper. The label ``autos" and ``RMS" indicate the method (either from products of auto-visibilities or the RMS of differenced visibilities) by which the curves of system temperatures are calculated. And the values of temperatures shown in labels are the average values over the band specified by the green regime. We see the results from two methods are consistent to $5\%$, though the curves from auto-correlations are far less scattered.}
    \label{fig:Tsys_comparison}
\end{figure*}

Another way to estimate $T_\text{sys}$---which we use in this paper---is to use auto-correlation visibilities, i.e., visibilities formed by correlating a single antenna's data with itself. The system temperature on a non-auto correlation baseline $\{p,q\}$ is then related to the geometric mean of the auto-correlation visibilities of the two constituent antennas as \citep{2015ApJ...801...51J}
\begin{equation}
\label{eq:auto_Tsys}
\sqrt{V(\{p,p\}) V(\{q,q\})} = \frac{2k_b \nu^2\Omega_p}{c^2} T_\text{sys,\{p,q\}}\,.
\end{equation}
In Figure \ref{fig:Tsys_comparison} we plot the system temperatures predicted using both methods for some HERA data. The lower scatter with the second method is why we recommend its usage.

The noise power spectrum $P_\text{N}$ correctly describes the error bars \emph{assuming that our instrument measures nothing but noise}. This may be a suitable approximation for noise-dominated delays. More generally, however, when a signal (be it foregrounds or systematics) exists, the cross terms of Equation~\eqref{eq:wedge} provide an additional contribution to the noise scatter/error bars.\footnote{We stress that this scatter/error is still due to instrumental noise and not the variance of the signal term. Even for a perfectly constant and known signal, the presence of the cross term alters the uncertainty, essentially having the signal term act as a multiplicative amplifier for noise fluctuations.} This term exists regardless of whether one's foreground mitigation strategy is based on subtraction or avoidance. In the former case, the foreground residuals after subtracting a model from data enter into the final expression; in the latter case, the whole foreground contribution is propagated as a systematic signal in the data. We show how to take this into account in Appendix \ref{ap:fg_dependent_var}, where we define $P_\text{SN}$ as
\begin{equation}
\label{eq:P_SN}
P_\text{SN}^2 \equiv \sqrt{2}\text{Re} (P_{\tilde{x}_1\tilde{x}_2}) P_\text{N} + P_\text{N}^2 \,
\end{equation}
which serves as a characterization of the error bars on the total sky emission, consistent with the form derived in \citet{kolopanis2019simplified}.
Here, $\text{Re} (P_{\tilde{x}_1\tilde{x}_2})$, the real part of power spectra formed from $\bm{x}_1$ and $\bm{x}_2$, serves as a stand-in for a signal-only power spectrum $P_\text{S}$ assuming that the signal dominates the noise (whether this ``signal" takes the form of foregrounds, systematics, or the cosmological signal).

Using real data helps us approximate the true $P_\text{S}$ when we do not possess good \emph{a priori} models. However, by using real data our estimate of the first term of Equation~\eqref{eq:P_SN} can in principle be negative because $\tilde{x}_1$ and $\tilde{x}_2$ contain different noise realizations. This can cause problems, since the signal-only power spectrum is expected to be non-negative. We thus enforce a hard prior on this term and set negative values of $\text{Re} (P_{\tilde{x}_1\tilde{x}_2})$ to zero. In this way $P_\text{SN}^2$ is always positive and the error bar $P_\text{SN}$ is at worst a conservative estimate. When we average power spectra with error bars, this conservatism leads to a substantial bias between $P_\text{SN}$ and $P_\text{N}$ in our final error estimates in the noise-dominated regime. This is due to $\text{Re} (P_{\tilde{x}_1\tilde{x}_2})$ in the first term of Equation \eqref{eq:P_SN} is empirical---and therefore contains noise---which effectively yields a double-counting of the noise-noise term in the variance. This double-counting does not result in an average bias if one does not enforce our prior, since in a noise-dominated regime $\text{Re} (P_{\tilde{x}_1\tilde{x}_2})$ has zero mean. Our prior ensures that $P_\text{SN} > P_\text{N}$. Despite this, we will show that Equation~\eqref{eq:P_SN} is a reasonable approximation over broad swaths of the power spectrum. Moreover, if we understand the statistics of noise fluctuations, one can simply predict---and correct for---the double-counting bias in $P_\text{SN}$. In the noise-dominated regime, $P_\text{N}$ characterizes the scatter in $\text{Re} (P_{\tilde{x}_1\tilde{x}_2})$. Thus one can estimate the expectation value of the extra noise contribution from the first term of Equation \eqref{eq:P_SN} by computing
\begin{align}
\label{eq:P_SN_bias}
& \sqrt{2} \langle \text{Re} (P_{\tilde{x}_1\tilde{x}_2}) \rangle  P_\text{N}\notag \\
= & \sqrt{2} \left[ \frac{1}{\sqrt{2\pi}P_\text{N}}\int_0^\infty y \exp{(-y^2/2P_\text{N}^2)} \text{d}y \right] P_\text{N} \notag \\
= & P_\text{N}^2/\sqrt{\pi} \,.
\end{align}  
The integral runs over only positive values since we are imposing a non-negative prior. Note that here where we have neglected any complicated window function effects in inserting the measured power spectrum, essentially assuming that all power is locally sourced at the delay where it is measured. In principle, these effects can be taken into account in a more general derivation within the quadratic estimator formalism, but we leave this for future work.

We see from Equation~\eqref{eq:P_SN_bias} that the excess of $P_\text{SN}$ above $P_\text{N}$ in the noise-dominated regime is proportional to $P_\text{N}$; thus, we can just subtract it from the initially computed $P_\text{SN}$. We then define a modified ``$P_\text{SN}$" free from the double-counting noise bias as\footnote{Here we derived the correction factor $\sqrt{1/\sqrt{\pi}+1} - 1 \approx 0.251$ assuming $\text{Re} (P_{\tilde{x}_1\tilde{x}_2})$ follows Gaussian distribution. This is appropriate assuming that enough power spectra formed from data at different times have been incoherently averaged together for the Central Limit Theorem to apply (we will examine this point further in Section~\ref{subsec:toymodel}). For a single snapshot in time, the measured power spectrum follows a Laplacian distribution (again, see Section~\ref{subsec:toymodel}) and the correction factor becomes $\sqrt{3/2}-1\approx 0.225$. Since the difference is small and in practice we operate in the Gaussianized regime anyway we use $\sqrt{1/\sqrt{\pi}+1} - 1$ in our definition.}
\begin{equation}
\label{eq:modified_P_SN}
\tilde{P}_\text{SN} \equiv P_\text{SN} - \left(\sqrt{1/\sqrt{\pi}+1} - 1\right)P_\text{N}\,.
\end{equation}
The reduction of double-counting noise bias in this way also holds where signal dominates over noise. Since $P_\text{N}$, $P_\text{SN}$, and $\tilde{P}_\text{SN}$ are all either power spectra or constructed from products of power spectra, we name this methodology of error estimation the ``Power Spectrum Method''.   

\subsection{Covariance Method} 
\label{subsec:cov_Method}
The quadratic estimator formalism leads to a natural way to write down an analytic form of error bars by propagating the input covariance matrices on visibilities into the output covariance matrices on bandpowers, which we name ``Covariance Method'' (see Appendix \ref{ap:covariance} for more details). Provided three set of matrices below containing the full frequency-frequency two-point correlation information of complex visibilities  
\begin{align}
\label{eq:covariance}
\bm{C}^{12}_{ij} \equiv & \langle \bm{x}_{1,i} \bm{x}_{2,j}^* \rangle \,, \notag\\
\bm{U}^{12}_{ij} \equiv & \langle \bm{x}_{1,i} \bm{x}_{2,j}\rangle  \,, \notag \\
\bm{G}^{12}_{ij} \equiv & \langle \bm{x}_{1,i}^* \bm{x}_{2,j}^*\rangle \,,
\end{align}  
the variance in the real part of $\hat{P}_\alpha$ is
\begin{align}
\label{eq:var_in_ps_real}
    &\phantom{=} \text{var} \left[\text{Re} (\hat{P}_\alpha) \right] \notag \\
    & = \frac{1}{4} 
    \Big\{\text{tr}\big[ (\bm{E}^{12,\alpha} \bm{U}^{22} \bm{E}^{21,\alpha*} \bm{G}^{11} + \bm{E}^{12,\alpha} \bm{C}^{21} \bm{E}^{12,\alpha} \bm{C}^{21}) \notag \\
    & +2\times (\bm{E}^{12,\alpha} \bm{U}^{21} \bm{E}^{12,\alpha *} \bm{G}^{21}+\bm{E}^{12,\alpha} \bm{C}^{22} \bm{E}^{21,\alpha} \bm{C}^{11})   \notag\\
    & + (\bm{E}^{21,\alpha} \bm{U}^{11} \bm{E}^{12,\alpha*} \bm{G}^{22} + \bm{E}^{21,\alpha} \bm{C}^{12} \bm{E}^{21,\alpha} \bm{C}^{12}) \big] \Big\}\,,
\end{align}
and the variance in the imaginary part of $\hat{P}_\alpha$ is
\begin{align}
\label{eq:var_in_ps_imag}
 &\phantom{=} \text{var} \left[\text{Im} (\hat{P}_\alpha) \right] \notag \\
   & = \frac{-1}{4} 
    \Big\{\text{tr}\big[ (\bm{E}^{12,\alpha} \bm{U}^{22} \bm{E}^{21,\alpha*} \bm{G}^{11} + \bm{E}^{12,\alpha} \bm{C}^{21} \bm{E}^{12,\alpha} \bm{C}^{21}) \notag \\
    & -2\times (\bm{E}^{12,\alpha} \bm{U}^{21} \bm{E}^{12,\alpha *} \bm{G}^{21}+\bm{E}^{12,\alpha} \bm{C}^{22} \bm{E}^{21,\alpha} \bm{C}^{11})   \notag\\
    & + (\bm{E}^{21,\alpha} \bm{U}^{11} \bm{E}^{12,\alpha*} \bm{G}^{22} + \bm{E}^{21,\alpha} \bm{C}^{12} \bm{E}^{21,\alpha} \bm{C}^{12}) \big] \Big\}\,,
\end{align}
To get the final error bar on power spectra, we should accurately model input covariance matrices on visibilities and propagate them into output covariance matrix on bandpowers. Generally, we assume that the input covariance matrices can be decomposed as $\bm{C} \equiv \bm{C}_\text{signal} + \bm{C}_\text{noise}$.

 Assuming the distributions of the real and imaginary parts of noise in visibilities are independently and identically distributed (IID) at the same frequency and are uncorrelated between different frequency channels, our expressions simplify considerably. With these assumptions, $\bm{C}_\text{noise}^{11}$ and $\bm{C}_\text{noise}^{22}$ are diagonal and $\bm{C}_\text{noise}^{12}$, $\bm{U}_\text{noise}^{11}$, $\bm{U}_\text{noise}^{22}$, $\bm{U}_\text{noise}^{12}$, $\bm{G}_\text{noise}^{11}$, $\bm{G}_\text{noise}^{22}$ and $\bm{G}_\text{noise}^{12}$ are all zero. Analogous to Equation~\eqref{eq:auto_Tsys}, one can estimate the diagonal terms of $\bm{C}_\text{noise}^{11}$ and $\bm{C}_\text{noise}^{22}$ using the amplitudes of auto-correlation visibilities. For a baseline $\{p,q\}$ composed by two antennas $p$ and $q$, its $\bm{C}_\text{noise}$ is 
\begin{align}
\label{eq:C_noise}
   \bm{C}^{\{p,q\}, \{p,q\}}_{\text{noise},ii}(t) \equiv & \phantom{-} \langle V_\text{noise}(\{p,q\},\nu_i,t) V_\text{noise}^*(\{p,q\},\nu_i,t) \rangle \notag \\
   \approx &  \phantom{-} \left|\frac{V(\{p,p\}, \nu_i,t)V(\{q,q\}, \nu_i,t)}{N_\text{nights} B \Delta t}\right| \,,
\end{align}
where $B\Delta t$ is the product of the channel bandwidth and the integration time, and $N_\text{nights}$ is the total number of nights of data analyzed from a drift scan telescope.

Inserting only $\bm{C}_\text{noise}$ for $\bm{C}$ in Equations \eqref{eq:var_in_ps_real} and \eqref{eq:var_in_ps_imag}, we have another estimate on the noise power variance as 
\begin{align}
\label{eq:QE-N}
 \text{var} \left[\text{Re} (\hat{P}_\alpha) \right] & = \text{var} \left[\text{Im} (\hat{P}_\alpha) \right]
 \notag \\ & = \frac{1}{2}
   \Big\{ \text{tr} \big[\bm{E}^{12,\alpha} \bm{C}_\text{noise}^{22} \bm{E}^{21,\alpha} \bm{C}_\text{noise}^{11}\big] \Big\} \notag \\
   & = \sigma_\text{QE-N}^2\,.
\end{align}
By taking the trace on the products of matrices, we have in fact taken a weighted average of covariance information over frequencies. The quantity $\sigma_\text{QE-N}$ should be equal to $P_\text{N}$ from the previous subsection, provided that in computing $T_\text{sys}$ using Equation \eqref{eq:P_N} we average over frequencies to obtain an effective $T_\text{sys}$ in the same way. In this way, we see that the analytic noise power spectrum essentially reduces to a special case of Equation \eqref{eq:QE-N}.      

Of course, the fully covariant treatment here also implicitly includes the signal-noise cross terms discussed in previous sections. Including both $\bm{C}_\text{signal}$ and $\bm{C}_\text{noise}$ in $\bm{C}$ gives
\begin{align}
\label{eq:QE-SN}
    \text{var} \left[\text{Re} (\hat{P}_\alpha) \right] 
    & = \text{var} \left[\text{Im} (\hat{P}_\alpha) \right]\notag \\
    & = \frac{1}{2} 
    \Big\{ \text{tr} \big[\bm{E}^{12,\alpha} \bm{C}_\text{noise}^{22} \bm{E}^{21,\alpha} \bm{C}_\text{noise}^{11} 
     \notag \\ 
    & \phantom{=} 
    + \bm{E}^{12,\alpha} \bm{C}_\text{signal}^{22} \bm{E}^{21,\alpha} \bm{C}_\text{noise}^{11}
     \notag \\ 
    & \phantom{=} 
     + \bm{E}^{12,\alpha} \bm{C}_\text{noise}^{22} \bm{E}^{21,\alpha} \bm{C}_\text{signal}^{11} \big]  \Big\} \notag \\
     & = \sigma_\text{QE-SN}^2 \,. 
\end{align}
Since we have assumed only $\bm{C}_\text{noise}^{11}$ and $\bm{C}_\text{noise}^{22}$ are non-zero, the extra signal-noise cross terms entering into the expression are just their couplings with the signal counterparts. For that last contribution, we estimate $ \bm{C}_\text{signal}$ as
\begin{align}
\label{eq:C_signal}
    \bm{C}_{\text{signal}, ij}^{11} = \bm{C}_{\text{signal}, ij}^{22} = \frac{1}{2}\left[\bm{x}_{1,i} \bm{x}_{2,j}^* + \bm{x}_{2,i} \bm{x}_{1,j}^*\right]\,.
\end{align}
Note that this way of modelling $\bm{C}_\text{signal}$ is Hermitian and noise-bias free when taking the ensemble average, but not positive definite. With a similar argument to $P_\text{SN}$ in subsection \ref{subsec:ps_method}, we enact a hard non-negative prior on $\bm{C}_\text{signal}$, where rows and columns containing negative diagonal elements are set to zero. This procedure can be shown to give signal-noise cross terms in Equation \eqref{eq:QE-SN} that are always non-negative.  However, this means that $\sigma_\text{QE-SN}$ suffers from the same double-counting noise bias with $P_\text{SN}$, and analogously we may construct a modified ``$\sigma_\text{QE-SN}$" which is also free from the bias as 
\begin{equation}
\label{eq:modified_QE-SN}
\tilde{\sigma}_\text{QE-SN} = \sigma_\text{QE-SN} - \left(\sqrt{1/\sqrt{\pi}+1} - 1\right)\sigma_\text{QE-N}\,.    
\end{equation}

Generally speaking, the power spectrum method of the previous subsection is a special case of the covariance method of this subsection. For example, if we estimate $P_\text{N}$ in a way that carefully accounts for the frequency dependence of $T_\text{sys}$, we should find that when we insert it into the expression for $P_\text{SN}$ that $P_\text{SN} = \sigma_\text{QE-SN}$. The covariance method has the advantage of providing off-diagonal covariances between different bandpowers in addition to variances.

\subsection{Summary}
The methods of error bar estimation introduced in this section can be categorized into two groups:
\begin{itemize}
    \item $\sigma_\text{bs}, P_\text{SN}, \sigma_\text{QE-SN}$: these estimate error bars on the total emission, including both contributions from signal-noise cross terms and noise-noise terms.  
    \item $P_\text{diff}$, $P_\text{N}, \sigma_\text{QE-N}$: these estimate the error bar in the limit of noise-dominated (or noise level), only including contributions from the noise-noise terms.
\end{itemize}
Before we jump into a quantitative discussion using the HERA power spectrum pipeline to compute these error bars in the next section, it is important to stress that there are other methods of error estimation that we do not cover in this paper. For example, LOFAR has used the Stokes V parameter as an estimator of noise level \citep{patil2017upper, 2019MNRAS.488.4271G, mertens2020improved} since the astrophysical sky is expected to exhibit only extremely weak circular polarization. However, reliably estimating Stokes V power requires more accurate polarization calibration solutions than that are currently available for HERA \citep{2019ApJ...882...58K}. Since one of our goals is to test our error estimation methods on HERA data, we will omit discussion of Stokes V techniques in this paper.

\section{Tests}
\label{sec:tests}
%%%
\begin{figure*}[!htbp]
\centering  
\includegraphics[width=\textwidth]{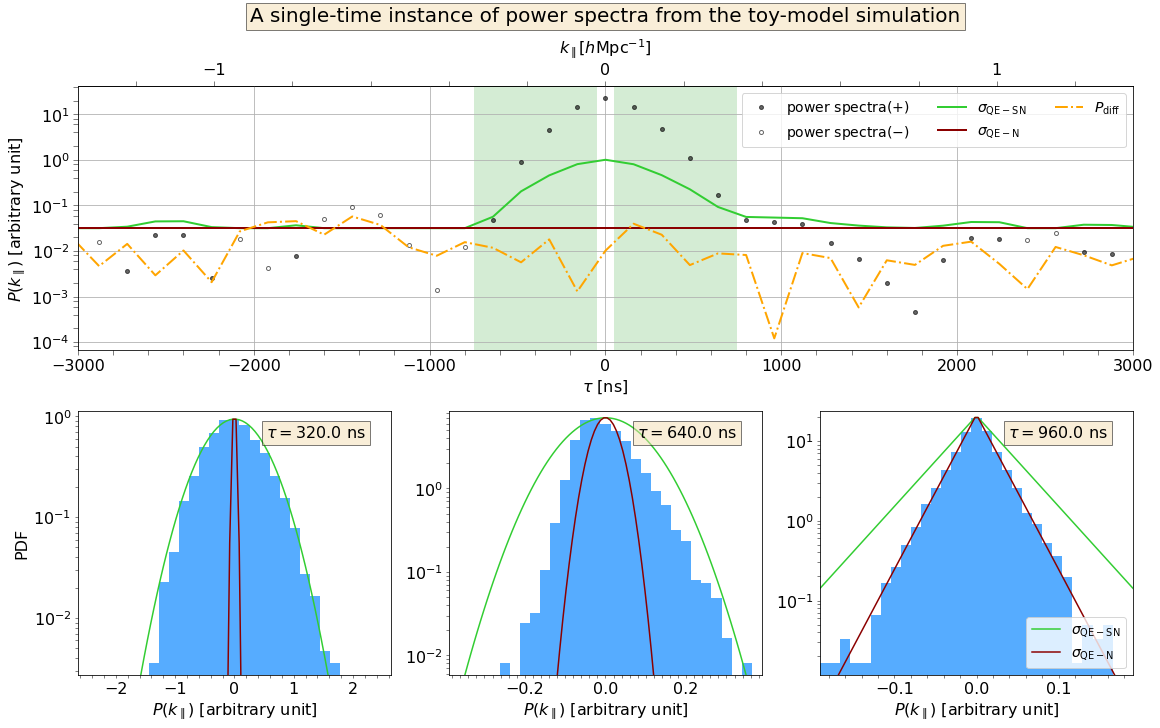}
 \caption{Error bars on single-baseline-pair power spectra at one timestamp from simulations described in Section \ref{subsec:toymodel}. \textit{Top}: We plot power spectra together with error bar types $P_\text{diff}, \sigma_\text{QE-SN}$ and $\sigma_\text{QE-N}$. The green shaded regime ranges from $\pm 50$ ns to $\pm 750$ ns, where the foreground power is dominant over the noise power. \textit{Bottom}: We plot histograms of bandpowers from $\sim 10000$ realizations at $\tau =  320.0\,\text{(strongly foreground-dominated regime)},\,640.0\,\text{(transition regime)}, \,960.0\,\text{(noise-dominated regime)}$ ns respectively, along with probability distribution function (PDF) curves predicted using the $\sigma_\text{QE-SN}$ and $\sigma_\text{QE-N}$ values at the same delay. At $\tau = 320.0, 640.0$ ns, the PDF takes a Gaussian form. At $\tau = 960.0$ ns, the PDF takes the form of a Laplacian. The $P(k_\parallel)$ values used in the histograms have been subtracted from the mean value of all realizations. We can see error bars are roughly comparable to each other in amplitudes in the noise-dominated regime. At $\tau = 320.0$, the envelope of the histogram matches exactly with the PDF using $\sigma_\text{QE-SN}$. At $\tau = 960.0$, the envelope of the histogram matches the PDF using $\sigma_\text{QE-N}$, while we see the PDF using $\sigma_\text{QE-SN}$ is broader. Therefore, using $\sigma_\text{QE-SN}$ will lead to a more conservative estimate of errors in this delay regime.} 
  \label{fig:toy_single_ps}
\end{figure*}

In this section, we quantitatively examine the error estimation methods introduced in Section \ref{sec:error}. We apply them to 21\,cm delay power spectra estimated from both simulated data and HERA Phase I data. We directly compare the relative amplitudes of the error bars predicted by each method, delay mode by delay mode. We also study how the error bars respond to systematics and foregrounds in different regimes of delay space. 

\begin{figure*}[!htbp]
\centering  
\includegraphics[width=\textwidth]{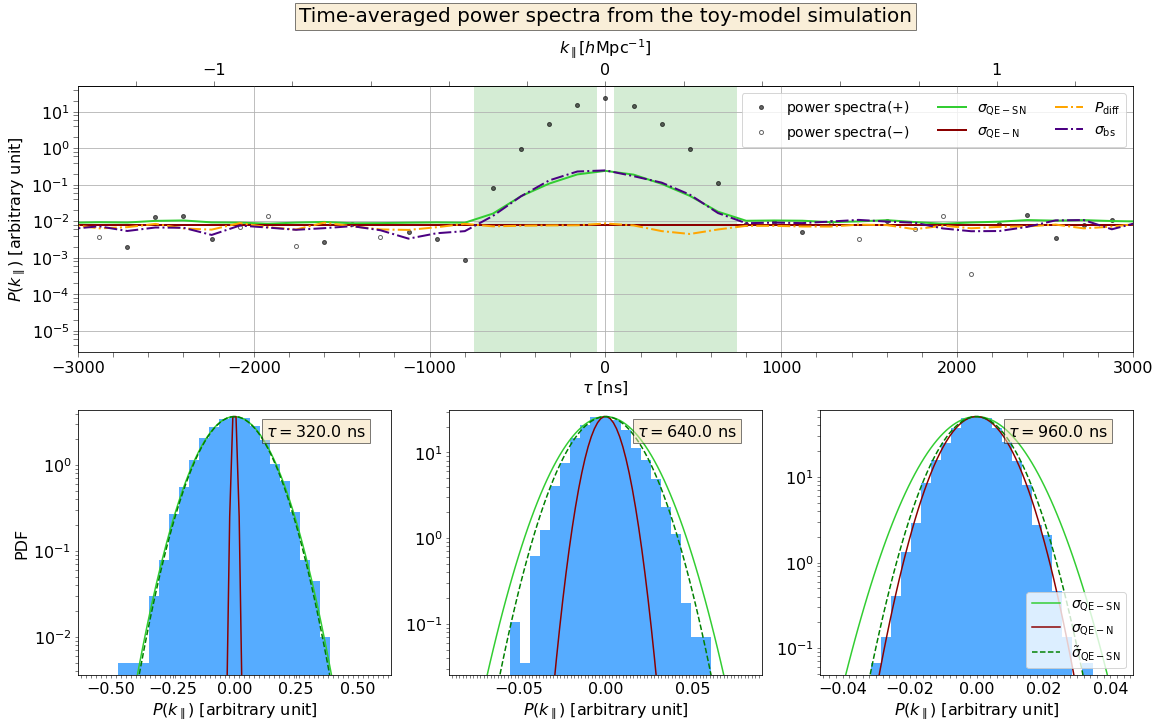}
 \caption{Error bars on time-averaged power spectra over 20 timestamps from simulations in Section \ref{subsec:toymodel}. The figure follows similar conventions to Figure \ref{fig:toy_single_ps}, except \textit{Top}: $\sigma_\text{bs}$ is added; \textit{Bottom}: All PDFs take the forms of Gaussian and the ones specified by $\tilde{\sigma}_\text{QE-SN}$ are appended. We observe good agreement between $\sigma_\text{bs}$ and $\sigma_\text{QE-SN}$ in the foreground-dominated regime, and the consistency of all types of labeled error bars in the noise-dominated regime. After the incoherent average, we see histograms at all delays become Gaussian. Additionally, $\tilde{\sigma}_\text{QE-SN}$ is clearly different from $\sigma_\text{QE-SN}$ where the signal is less dominant. Especially at $\tau = 960.0$ ns, the PDF using $\tilde{\sigma}_\text{QE-SN}$ is closer to the exact noise-dominated version using $\sigma_\text{QE-N}$.}
  \label{fig:toy_average_ps}
\end{figure*}

\subsection{Simulations from a Toy Model}
\label{subsec:toymodel}
We start with simulations from a toy model. This allows us to generate a large number of realizations, with which we can numerically test the validity of our error bars in the ensemble-averaged limit. Our simulated visibilities include only the foregrounds and noise. For the foreground portion of the visibilities we draw a random visibility from a frequency-frequency covariance matrix of the form $\bm{C}_{ij} = A \exp{[-(\nu_i-\nu_j)^2/l^2]}$, where $A$ and $l$ characterize the amplitude and correlation length of the foreground signal, respectively. The adopted covariance model creates smoothly varying functions in frequency space, which is roughly in accordance with the relatively flat spectral structure of real foregrounds. Here we simulate visibilities on two redundant baselines for 20 consecutive timestamps. We set $A=25$ and $l=5\text{MHz}$, and the foreground visibilities are kept the same on each baseline and over all timestamps. The noise components of the visibilities on each baseline at each timestamp are independently drawn from the same white Gaussian distribution $\mathcal{N}(0, \sigma^2=1)$. We produce $\sim$ 10000 realizations of such visibilities and then use \texttt{hera\_pspec} code\footnote{\url{https://github.com/HERA-Team/hera_pspec}} to estimate the delay power spectra and to compute the error bars discussed previously.

In Figure~\ref{fig:toy_single_ps}, we plot power spectra together with a few of the error bar types computed from one timestamp of data from the simulations. We compute $P_\text{diff}$ by differencing visibilities between the one timestamp and the next. We use Equation \eqref{eq:QE-N} and \eqref{eq:QE-SN} to calculate error bars of the ``covariance method", while we evaluate $\bm{C}_\text{noise}$ using the exact covariance matrix from which noise visibilities are drawn, since we did not simulate visibilities on auto-correlation baselines. In the top panel of Figure \ref{fig:toy_single_ps}, the green shaded regime (which ranges from $\pm 50\,\textrm{ns}$ to $\pm 750\,\textrm{ns}$) is where the foreground power is dominant over the noise power. We see that $P_\text{diff}$ and $\sigma_\text{QE-N}$ are insensitive to the foreground power in this regime, and when moving to higher delays, the noise levels characterized by $P_\text{diff}, \sigma_\text{QE-N}$, and $ \sigma_\text{QE-SN}$ are very close to one another. Compared to the other two, $P_\text{diff}$ shows much more scatter from delay to delay since it is a more empirical estimation of noise based on examining what amounts to noise \emph{realizations}. Notice also that as expected by construction, the $\sigma_\text{QE-SN}$ curve always lies above $\sigma_\text{QE-N}$, due to the fact we enforce a zero clipping on the signal-noise cross term.

In the bottom panel of Figure \ref{fig:toy_single_ps}, we plot histograms of power spectra at three delays ($\tau = 320.0, 640.0$ and $960.0\,\textrm{ns}$) by accumulating data points from  $\sim10000$ realizations. 
The results here are therefore representative of ensemble-averaged expectations. At each delay, we also plot theoretical predictions for the probability distribution functions (PDFs). Precisely what form these PDFs take will depend on the delay. In the low-delay regime, Equation~\eqref{eq:wedge} shows the variation comes from single powers of visibility noise, which we assume is Gaussian. (Recall that we are not modelling the signal as a random field, in the sense that it does not participate in our ensemble average.) The result is a Gaussian PDF. At high delays Equation~\eqref{eq:wedge} shows that the power spectrum is the cross-multiplication of two independent realization of noise. The resulting PDF is a Laplacian. Both of these distributions take one free parameter (the standard deviation of power) and we show predictions where this standard deviation is specified by $\sigma_\text{QE-SN}$ and $ \sigma_\text{QE-N}$. At $\tau = 320.0$ and $640.0\,\textrm{ns}$, we plot Gaussian reference PDFs.  At $\tau = 960.0\,\textrm{ns}$, we plot a Laplacian reference PDF. We see at $\tau = 320.0\,\textrm{ns}$, where foreground power is overwhelmingly dominant, the shape of the histogram is indeed Gaussian-like, and its envelope matches the PDF curves using $\sigma_\text{QE-SN}$. At $\tau = 960.0$ where noise is dominant, the shape of the histogram is indeed Laplacian-like, and its envelope matches the PDF curves using $\sigma_\text{QE-N}$ (since $\sigma_\text{QE-N}$ does not suffer from the conservatism of $\sigma_\text{QE-SN}$ discussed in Section~\ref{subsec:ps_method}). With $\tau = 640.0\,\textrm{ns}$ we have a transition case between the two extremes. The distribution of power spectra will be skewed since neither the signal nor the noise dominates in this occasion (for a mathematical proof of the skewness see Appendix \ref{ap:skewness_intermediate}). The histogram does not match the PDF predicted by either standard deviation, but note from the widths of the PDFs that an error bar given by $\sigma_\text{QE-SN}$ is a conservative error, as we designed it to be.

In Figure \ref{fig:toy_average_ps}, we present the same types of error bars plus a bootstrapped one on power spectra which were formed by incoherently averaging over 20 timestamps. We see in the green regime that $\sigma_\text{bs}$ agrees with $\sigma_\text{QE-SN}$. All the different kinds of error bars agree well with each other in the noise dominated regime, and with the extra time averaging step (compared to Figure~\ref{fig:toy_single_ps}) $P_\text{diff}$ exhibits less scatter. Again, we plot histograms of the averaged power spectra from Monte-Carlo simulations against Gaussian PDF curves at $\tau = 320.0, 640.0$ and $960.0\,\textrm{ns}$. One feature to note from the histogram is that each distribution has become nearly Gaussian. This is simply due to the Central Limit Theorem as power spectra are averaged together incoherently. In addition to $\sigma_\text{QE-SN}$ and $\sigma_\text{QE-N}$, we also plot the PDFs using $\tilde{\sigma}_\text{QE-SN}$ which eliminates the double-counting bias in $\sigma_\text{QE-SN}$. It is as expected that the PDF using $\tilde{\sigma}_\text{QE-SN}$ is more close to the one using $\sigma_\text{QE-N}$ at the noise-dominated delay mode.   

\subsection{Application to HERA Phase I Data}  
\label{subsec:HERA_Data}
\begin{figure*}[!htbp]
\centering  
\includegraphics[width=0.9\textwidth]{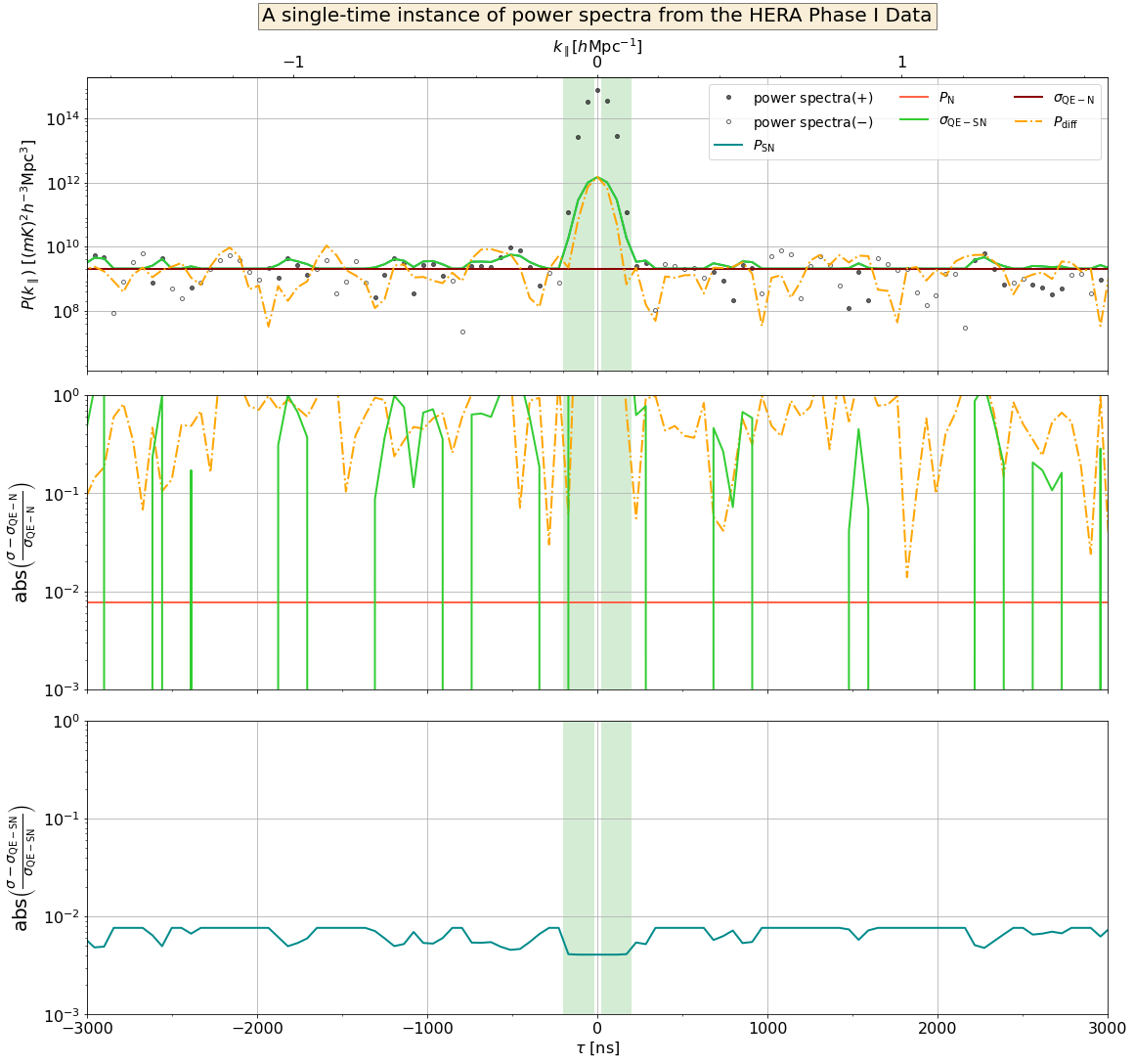}
\caption{Error bars on single-baseline-pair power spectra at one timestamp from HERA Phase I data. The visibilities are selected from a band spanning 150.3 to 167.8 MHz. \textit{Top}: Power spectra with error bars. The green shaded regime ranging from $\pm 20$ ns to $\pm 200$ ns is expected to be foreground dominated. \textit{Middle}: Absolute relative difference between selected error bars with $\sigma_\text{QE-N}$. \textit{Bottom}: Absolute relative difference between selected error bars with $\sigma_\text{QE-SN}$. We see numerically that $P_\text{SN}$ differs from $\sigma_\text{QE-SN}$ by less than 1\% and that the same is true for $P_\text{N}$ and $\sigma_\text{QE-N}$. }
\label{fig:IDR2_2_LPX_single_ps}
\end{figure*}

\begin{figure*}[!htbp]
\centering  
\includegraphics[width=0.9\textwidth]{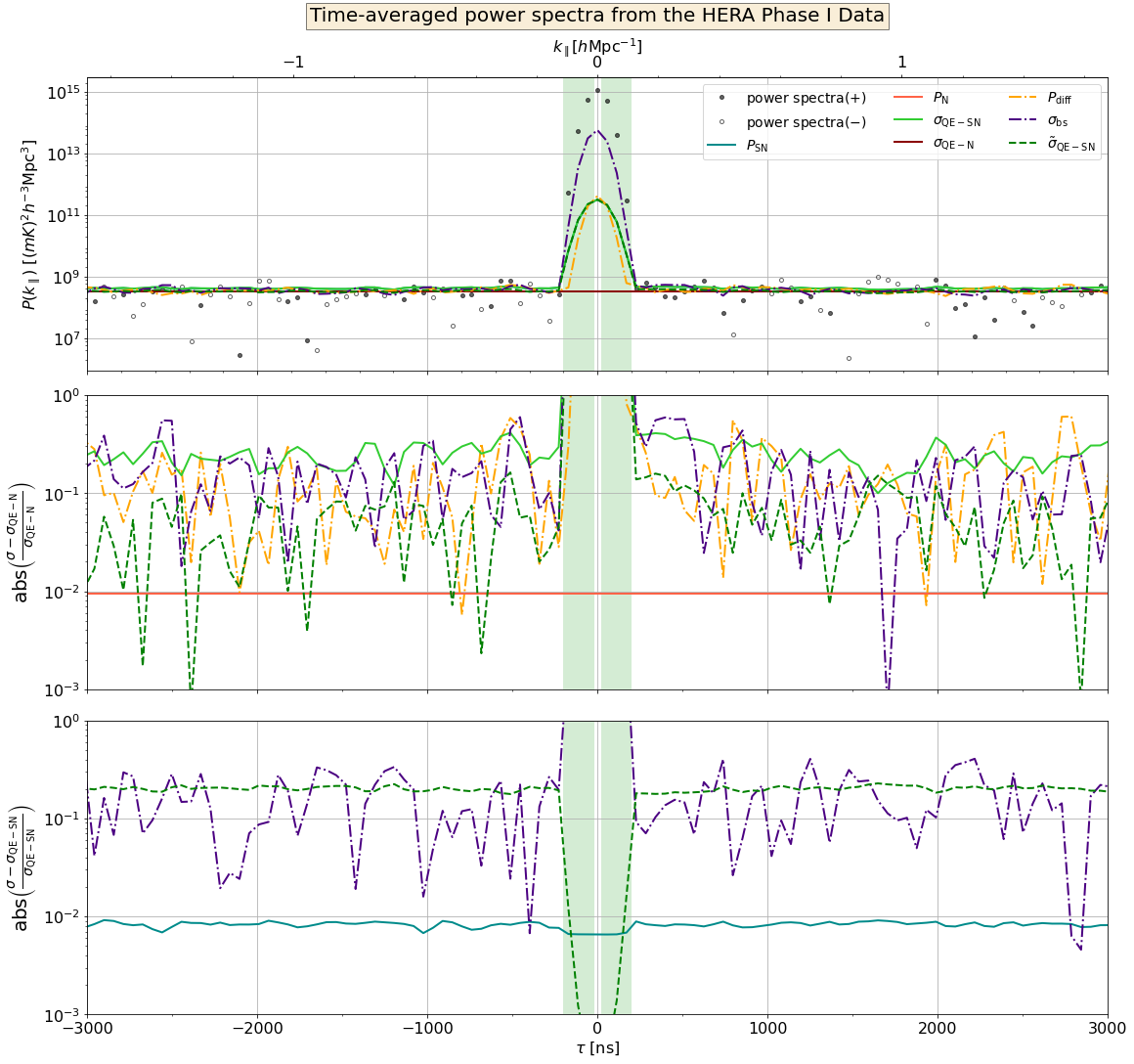}
\caption{Error bars on single-baseline-pair power spectra incoherently averaged over 30 time samples from the same slice of HERA Phase I data as Figure \ref{fig:IDR2_2_LPX_single_ps}. Our plotting conventions also follow those of Figure \ref{fig:IDR2_2_LPX_single_ps} for other conventions. We add results from $\tilde{\sigma}_\text{QE-SN}$ in each panel. In the center panel we see the relative difference between $\tilde{\sigma}_\text{QE-SN}$ and $\sigma_\text{QE-N}$ drops remarkably from $\sim 30\%$ to a few percent compared to the $\sigma_\text{QE-SN}$, demonstrating the effectiveness of our noise-double-counting bias removal. On the other hand, in the bottom panel we see that going from $\sigma_\text{QE-SN}$ to $\tilde{\sigma}_\text{QE-SN}$ results in significant changes only at the noise-dominated delays, and thus there one can always elect to use $\tilde{\sigma}_\text{QE-SN}$ even in foreground-dominated regimes. }
\label{fig:IDR2_2_LPX_average_ps}
\end{figure*}

\begin{figure*}[!htbp]
\centering  
\includegraphics[width=\textwidth]{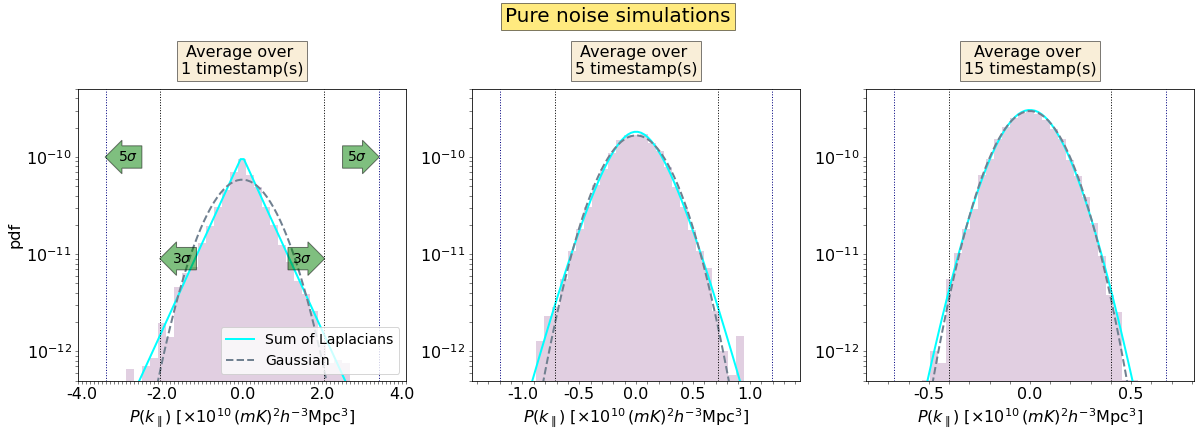}
 \caption{We plot the histograms of incoherently averaged power spectra over certain timestamps from pure noise simulations. The histogram in each column contains $\sim 10000$ data points. We compute $\sigma_\text{QE-N}$ and refer to Equation \eqref{eq:SumOfLaplace} to evaluate the ``Sum of Laplacians" PDF. Data points have been subtracted from the mean over all realizations. We also plot the equivalent Gaussian PDF with the same variance as the ``Sum of Laplacians" PDF. The green arrows point to the dotted vertical lines representing ``$3\sigma$" and ``$5\sigma$", where $\sigma$ is the square root of the variance of the predicted PDF. We see the envelopes of the histograms match the PDFs predicted using \eqref{eq:SumOfLaplace} very well. As a check, the fractions of outliers beyond $3\sigma$ in each histogram are $(1.27\%, 0.57\%, 0.25\%)$, while the corresponding values from the predicted PDFs are $(1.34\%, 0.58\%, 0.22\%)$---a very close agreement. And with more time samples to be incoherently averaged, the shape of the histogram becomes increasingly Gaussian, which is a consequence of the central-limit theorem. As expected, we also see the distribution get narrower with more samples averaged together. }
  \label{fig:null_test_noise}
\end{figure*}

\begin{figure*}[!htbp]
\centering 
\includegraphics[width=\textwidth]{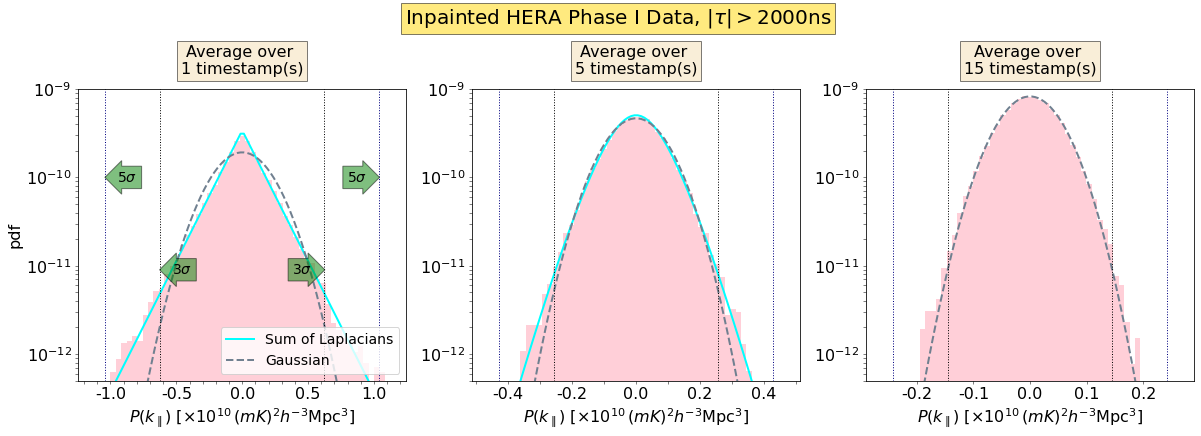}
\includegraphics[width=\textwidth]{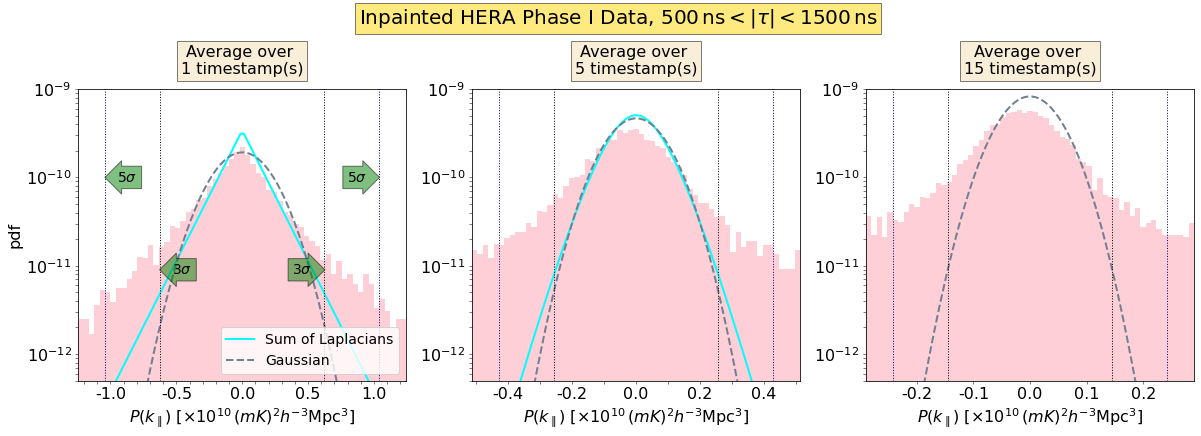}
 \caption{Histograms of power spectra at noise-like modes from the same HERA Phase I data used in Figure \ref{fig:IDR2_2_LPX_single_ps} and \ref{fig:IDR2_2_LPX_average_ps}, including RFI gap inpainting, but without the removal of systematics. The data points are accumulated from power spectra at the same delays from different redundant baseline-pairs. Because their noise levels may differ, they are first normalized by dividing out their corresponding $\sigma_\text{QE-N}$ and then having the mean of all data points subtracted off. In this way we have a uniform $\sigma_\text{QE-N}$ for all points, and we use Equation \eqref{eq:SumOfLaplace} to compute the ``Sum of Laplacians" PDF. Refer to Figure \ref{fig:null_test_noise} for other plotting conventions. \textit{Top}: histograms from power spectra at all delays larger than 2000 ns, where there are $\sim$ 27000 points in each column. \textit{Bottom}: histograms from power spectra at delays between 500 and 1500 ns, where there are $\sim 9000$ points in each column. As a check, in the top panel, the fractions of outliers beyond $3\sigma$ in each histogram are $(1.49\%, 0.65\%, 0.40\%)$, which are close to the corresponding values from the predicted PDFs $(1.36\%, 0.57\%, 0.24\%)$. In the bottom panel, the fractions of outliers beyond $3\sigma$ in each histogram are $(7.95\%, 10.70\%, 11.46\%)$, which greatly exceed corresponding values from the predicted PDFs $(1.36\%, 0.57\%, 0.24\%)$. }
  \label{fig:inpainted_only}
\end{figure*}

\begin{figure*}[!htbp]
\centering  
\includegraphics[width=\textwidth]{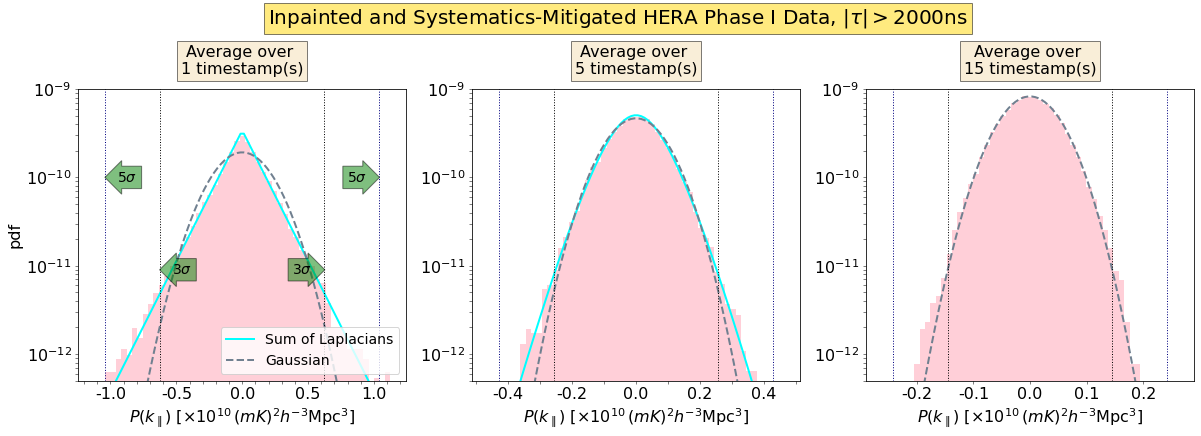}
\includegraphics[width=\textwidth]{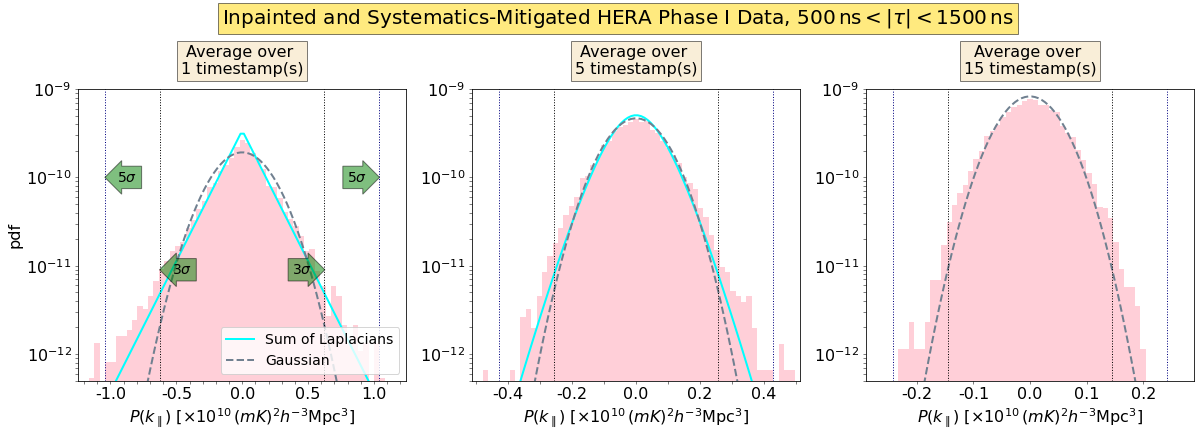}
 \caption{Histograms of power spectra at noise-like modes from inpainted and systematics-mitigated HERA Phase I data. The power spectra used here come from  exactly the same data set as Figure \ref{fig:IDR2_2_LPX_single_ps} and \ref{fig:IDR2_2_LPX_average_ps}. As a check, in the top panel the fractions of outliers beyond $3\sigma$ in each histogram are $(1.48\%, 0.63\%, 0.39\%)$, which are close to the corresponding values from the predicted PDFs $(1.36\%, 0.57\%, 0.24\%)$. And in the bottom panel, the fractions of outliers beyond $3\sigma$ in each histogram are $(2.19\%, 1.32\%, 0.80\%)$, which slightly exceed the corresponding values from the predicted PDFs $(1.36\%, 0.57\%, 0.24\%)$, but at a much lower level than the disagreement seen in Figure~\ref{fig:inpainted_only}.}
  \label{fig:null_test_IDR2.2}
\end{figure*}
The HERA Phase I data used for analysis in this paper consists of 18 observing nights taken in the Karoo Desert, South Africa from December 10th to 28th, 2017. The HERA array consisted of $\sim40$ functional antennas during observations, which were taken across a $100$ to $200\,\textrm{MHz}$ band comprised of 1024 channels and dual polarization ``X" and ``Y" feeds. [See Table 1 of \citet{kern2020abscal} for more details on the array and correlator specifications during the observations.] The data used in this work were first preprocessed with the HERA analysis pipeline (internally called H1C IDR2.2\footnote{\url{http://reionization.org/manual_uploads/HERA069_IDR2.2_Memo_v3.html}}). This includes automated metric evaluation and data flagging for faulty antennas and radio frequency interference (RFI).
In addition, the data are redundantly calibrated \citep{dillon2020redcal}, absolutely calibrated \citep{kern2020abscal}, binned and averaged across observing nights, in-painted over RFI gaps in frequency and then treated for known instrumental systematics \citep{kern2020mitigating}. 

We pick a slice of HERA Phase I visibilities taken from a 14.6-m redundant baseline group during an LST range of $5.75$ to $6.10$ hours. The visibilities in each timestamp are integrated over $\sim 10$ seconds. We select visibilities falling within a $150.3$ to $167.8\,\textrm{MHz}$ band to compute power spectra. We use pseudo-Stokes I visibilities $V_\text{pI}$, which are constructed by combining the visibilities from a cross correlation of two X feeds (``XX") and a cross-correlation two Y feeds (``YY") as follows:
\begin{equation}
\label{eq:stokes_pI}
V_\text{pI} = \frac{1}{2}\left(V_\text{XX} + V_\text{YY} \right)\,.
\end{equation}
In forming the delay power spectra we cross correlate visibilities from different baselines (e.g., $\bm{b}_1$-$\bm{b}_2$, $\bm{b}_1$-$\bm{b}_3$, $\bm{b}_2$-$\bm{b}_3$, etc.) and between odd and even timestamps (e.g., $t_1$-$t_2$, $t_3$-$t_4$, $t_5$-$t_6$, etc.) to form delay power spectra. In this way, we obtain power spectra on 253 baseline-pairs at 30 timestamps.

We show the power spectra from one baseline-pair at one timestamp in Figure \ref{fig:IDR2_2_LPX_single_ps}, together with error bar types $P_\text{diff}, \sigma_\text{QE-SN}, \sigma_\text{QE-N}, P_\text{SN}$, and $P_\text{N}$. The $P_\text{diff}$ errors are computed from time-differenced visibilities, e.g., for power spectra at the cross timestamp $t_1-t_2$ we form $V_\text{diff} \propto V(t_2) - V(t_1)$ and then we cross multiply $V_\text{diff}$ from two different baselines to obtain the corresponding $P_\text{diff}$ for that baseline pair. We calculate $\sigma_\text{QE-SN}$ and $\sigma_\text{QE-N}$ using Equations~\eqref{eq:QE-SN} and \eqref{eq:QE-N} with $\mathbf{C}_\text{signal}$ and $\mathbf{C}_\text{noise}$ specified by Equation \eqref{eq:C_signal} and \eqref{eq:C_noise}. Equations~\eqref{eq:P_SN} and \eqref{eq:P_N} give the expressions for $P_\text{SN}$ and $P_\text{N}$. See \texttt{hera\_pspec} for detailed implementation.

In the top panel of Figure \ref{fig:IDR2_2_LPX_single_ps}, we see all error bars agree well with each other in the noise-dominated regime (the red curve for $P_\textrm{N}$ is almost exactly underneath the brown curve for $\sigma_\text{QE-N}$, making the former difficult to see; the same is true for the teal curve for $P_\textrm{SN}$ versus the bright green curve for $\sigma_\text{QE-SN}$). The green shaded regime ranging from $\pm 20\,\textrm{ns}$ to $\pm 200\,\textrm{ns}$ is where foregrounds are expected to dominate. Here we see that  $P_\text{diff}$ also responds to the foreground power, similar to $P_\text{SN}$ and $\sigma_\text{QE-SN}$. This tells us that the time-differenced visibilities contain non-negligible foreground residuals, which is not surprising since the sky is expected to evolve non-negligibly over the $\sim$ 10 seconds of difference between our time samples.

In Section \ref{sec:error}, we argued that the ``covariance method" and the ``power spectrum method" should be equivalent to each other. In the middle and bottom panels of Figure \ref{fig:IDR2_2_LPX_single_ps}, we compute the relative difference in magnitudes between error bars, setting $\sigma_\text{QE-SN}$ and $\sigma_\text{QE-N}$ as the benchmarks respectively. We see that $P_\text{SN}$ differs from $\sigma_\text{QE-SN}$ and $P_\text{N}$ from $\sigma_\text{QE-N}$ by less than 1\%, so they are essentially equivalent in our pipeline. On the other hand, $P_\text{diff}$ can differ from $\sigma_\text{QE-N}$ at more than the 10\%-level due to the fact that it is highly scattered. Note that $\sigma_\text{QE-SN}$ and $P_\text{SN}$ are also scattered at some delays, whereas they are equal to $\sigma_\text{QE-N}$ and $P_\text{N}$ at other delays. This is due to our imposition of a non-negative prior on the signal-noise cross term.       

In Figure \ref{fig:IDR2_2_LPX_average_ps}, we show the power spectra with error bars on the same baseline-pair as Figure \ref{fig:IDR2_2_LPX_single_ps},  but with the further step of incoherently averaging over 30 time samples. We still see that all error bars (with bootstrap errors $\sigma_\text{bs}$ added) agree well in the noise-dominated regime. At low delays, $\sigma_\text{bs}$ peaks at an even higher value than $\sigma_\text{QE-SN}$. This is because the sky is not unchanged over different timestamps, so the bootstrapped error bars over time samples are inflated. After incoherently averaging, we still see $P_\text{SN}$ differing from $\sigma_\text{QE-SN}$ and $P_\text{N}$ differing from $\sigma_\text{QE-N}$ by less than 1\%. On the other hand, $P_\text{diff}$ and $\sigma_\text{bs}$ differ from $\sigma_\text{QE-N}$ at roughly the 10\% level in the noise-dominated regime. We also see that in the limit of noise domination, $\sigma_\text{QE-SN}$ has a relative bias over $\sigma_\text{QE-SN}$ by about 30\%. Therefore, using $\sigma_\text{QE-SN}$ or $P_\text{SN}$ leads to a conservative estimate of one's errors, as we expected. For comparing, we also plot results of $\tilde{\sigma}_\text{QE-SN}$, which eliminates the double-counting noise bias in $\sigma_\text{QE-SN}$. The relative difference between $\tilde{\sigma}_\text{QE-SN}$ and $\sigma_\text{QE-N}$ is reduced to a few percents in the noise-dominated regime. While $\tilde{\sigma}_\text{QE-SN}$ is not significantly modified from $\sigma_\text{QE-SN}$ in the foreground-dominated regime. Thus if we want a compromise on reflecting the properties of the signal-noise cross term while not introducing noise bias, $\tilde{\sigma}_\text{QE-SN}$ might be our choice.   

What we have established so far is the \emph{relative} agreement (or lack thereof) between different types of error bars in different regimes. However, we have not yet established the \emph{absolute} validity of these error bars on real data (i.e., we have not ruled out the possibility that they are all incorrect in the same way). For simulated power spectra we were able to compare the Monte-Carlo histograms with the PDF curves predicted from the error bars. The good match between the two gave us confidence in applying our error estimation methods. Might we perform similar analyses for power spectra from real data? Unfortunately, in real observations we only have one realization of the sky so that we cannot reach ensemble average limit by accumulating data points from a large number of realizations. Also, unlike simulated data with understood statistics, real data will contain systematics that make their statistics more complicated and difficult to understand (although this may change as the field of $21\,\textrm{cm}$ cosmology continues to mature).

For now, we may partially achieve our goal by checking the distributions of noise-like modes in our power spectra of real data. The noise-like modes refer to power spectra at higher delays where noise power is thought to be dominant and systematics are negligible. As we discussed in Section \ref{sec:error}, we expect the noise visibilities to be Gaussian-distributed. This makes it possible to analytically compute the resultant statistics of power spectra. In Appendix \ref{ap:pdf}, we derive the mathematical form of the PDF of incoherently averaged noise-dominated power spectra. The final result, Equation \eqref{eq:SumOfLaplace}, shows that the correct PDF is a weighted sum of a series of Laplacian distributions. As a numeric test of the derivation, we produce Monte-Carlo histograms of incoherently averaged power spectra from pure Gaussian noise visibilities with an increasing number of averaged samples in Figure \ref{fig:null_test_noise}. We generate $\sim 10000$ realizations of power spectra with multiple time samples, and evaluate the power spectra at a single timestamp, as well as what it would be if incoherently averaged over 5  or 15 timestamps. For realizations at each time sample, we can calculate the error bar $\sigma_\text{QE-N}$ of the power spectra and substitute them into Equation \eqref{eq:SumOfLaplace}. It is clear that the predicted PDF matches the envelope of the histograms and that the shape of the histograms of averaged power spectra become increasingly Gaussian when averaging is over more timestamps. This is again a result of the Central Limit Theorem.

Confronting our results with real data, we use the power spectra from the same HERA Phase I data set as Figures \ref{fig:IDR2_2_LPX_single_ps} and \ref{fig:IDR2_2_LPX_average_ps} to generate the histograms. To accumulate sufficient data points for a histogram, we view all noise-like modes in power spectra over different redundant baseline-pairs as independent realizations. And we carry out the incoherent average over the time axis. Because the noise level at different baseline pairs may differ, all power spectra are first normalized by being divided over their corresponding $\sigma_\text{QE-N}$ and then subtracted from the mean of all data points. After the normalization, we have a uniform error bar $\sigma_\text{QE-N}$ for all data points at each time sample. We then make histograms and compare their envelopes with the PDF of ``Sum of Laplacians" predicted using Equation \eqref{eq:SumOfLaplace}. 

Before we jump to the results, we first take a look at the data set which includes RFI gap inpainting but without the removal of systematics. For histograms drawn in Figure \ref{fig:inpainted_only}, we evaluate the distributions of power spectra at delays larger than $2000\,\textrm{ns}$ and at delays between $500$ and $1500\,\textrm{ns}$, respectively. In the former case, we see the shape of histograms are perfectly consistent with the predicted PDF, and the distributions become more Gaussian and narrower with increasing number of averaged samples, similar to what we saw in Figure \ref{fig:null_test_noise}. While in the latter case, we observe the histograms are flattened and much wider compared to the predicted PDF and there exist evidently hefty wings on either ends. Numerically, the fractions of outliers beyond $3\sigma$ in each histogram are $(7.95\%, 10.70\%, 11.46\%)$, which greatly exceed corresponding values from predicted PDFs $(1.36\%, 0.57\%, 0.24\%)$. This is a remarkable proof that significant systematics exist at lower delays in inpainted only data, as we expect. 

We produce histograms for the systematics-removed data, as we used for Figures \ref{fig:IDR2_2_LPX_single_ps} and \ref{fig:IDR2_2_LPX_average_ps}, in Figure \ref{fig:null_test_IDR2.2}. At delays larger than $2000\,\textrm{ns}$, we still see a good match between the Monte-Carlo histograms with the predicted PDFs. While at delays between $500$ and $1500\,\textrm{ns}$, we see the deviations between histograms and PDFs are highly suppressed, compared to Figure \ref{fig:inpainted_only}. This is not surprising since we have exerted systematics removal. Though there is still a little excess above PDFs in histograms on far ends, this does not substantially affect the error bars that one might quote on a power spectrum measurement (which serve as a summary statistic for the main bulk of the PDF rather than its wings). However, such deviations are worth keeping an eye on, especially when performing rigorous jackknife or null tests in an attempt to understand the systematics in one's instrument. As noted above, the excessive wings of the histograms in the bottom panel of Figure \ref{fig:inpainted_only} can serve as a diagnostic tool for systematics that lead to deviations from Gaussian noise-like visibilities. They may also be used to investigate the related question of how instrumental systematics (e.g., \citealt{kern2019mitigating, kern2020mitigating}) might affect the validity of one's error bars. Readers should interpret Figure \ref{fig:inpainted_only} and \ref{fig:null_test_IDR2.2} as a quality check of HERA Phase I data, which shows the power spectra at high delays ($> 2000$ ns) and at middle delays (500-1500 ns) after systematics mitigation are close to the predicted behaviors of Gaussian noise visibilities. Thus $\sigma_\text{QE-N}$ (along with other equivalent methods) validates itself a successful tool to characterize the noise statistics in real data. However, we will still quote $\tilde{\sigma}_\text{QE-SN}$ as a more robust error bar on reporting EoR upper limits at those delays. One should be aware that not all systematics can be cleanly corrected for, which mean that in principle the statistics can be much more complicated than the simple Gaussian distribution shown here. Along this theme, we urge readers to always perform consistency checks on the data, including but not limited to the ones we have performed here.

\section{Discussion}
\label{sec:discussion}
%%%
In previous sections, we have examined a number of different methods for assigning error bars to a HERA power spectrum. Here, we perform a comparison of the different types of error bars, highlighting the advantages and disadvantages of each.

We first consider the error bars using the ``covariance method" ($\sigma_\text{QE-N}$ and $\sigma_\text{QE-SN}$) to those computed using the ``power spectrum method" ($P_\text{N}$ and $P_\text{SN}$).
\begin{itemize}
    \item The ``covariance method" error bars analytically take the covariance of the input visibilities and propagate them through to the output covariance of the bandpowers, via general formulae given by Equations \eqref{eq:var_in_ps_real} and \eqref{eq:var_in_ps_imag}. There are two weaknesses to this approach. First, the output errors will only be as good as the modeling of the input covariances. This modeling is particularly difficult for foregrounds and systematics, which can have statistical properties that are not entirely understood. In this paper, we adopt a strategy where we view systematics as non-random, and empirically estimate them from the real data. The other weakness of our ``covariance method" is that our derivations rely on Gaussianity (Indeed, it would be strange for this method to only require an input \emph{covariance}---a two-point function---if it were capable of capturing the effects of non-Gaussianity). This assumption will also be violated by foregrounds and systematics as well the cosmological signal (which is an effect that was modeled in \citealt{mondal2015statistics, 2017MNRAS.464.2992M, 2019MNRAS.487.4951S}).
    
    Sidestepping these modeling restrictions on the ``covariance method" are the noise-dominated bandpowers at high delays. In this regime, we use an input covariance matrix that is $C_\text{noise}$ that is diagonal, with the diagonal elements set by the auto-correlation visibilities as Equation \eqref{eq:C_noise}. The resulting error bars we call $\sigma_\text{QE-N}$ (see Table~\ref{tab:intro} for a reminder of our notation). These error bars are confirmed by tests on simulations and real data in Figure \ref{fig:null_test_noise} and Figure \ref{fig:null_test_IDR2.2}, which verify that the error bars do properly account for the spread seen in an ensemble of Monte Carlo simulations. Further bolstering our confidence in using the ``covariance method" are their agreement with other error metrics at our disposal. Figures \ref{fig:IDR2_2_LPX_single_ps} and \ref{fig:IDR2_2_LPX_average_ps} show that in the noise-dominated regime, the error bars using the ``covariance method" are in excellent agreement with the bootstrap errors $\sigma_\text{bs}$, error bars using the `power spectrum method', and the power spectrum of differenced data $P_\text{diff}$.
    
    \item The agreement between these different error estimation methods raises the question of why one might favour the ``covariance method" over others. Consider first a comparison between $\sigma_\text{QE-N}$ and $P_\text{N}$ from the ``power spectrum method". These two methods are in fact quite similar, because $P_\text{N}$ is also an analytically propagated measurement of error, as one can see for instance in the derivation of \citet{ZFH2004}. The difference is one of generality, whether in the inputs, the intermediate steps, and the outputs. On the input side, $P_\text{N}$ assumes uncorrelated noise between visibilities whose amplitude is governed by the radiometer equation; $\sigma_\text{QE-N}$ can accept an arbitrary input covariance (even though in our tests we take it to be diagonal). During the actual propagation of errors, the derivation of $P_\text{N}$ assumes that fluctuations in $uv\nu$ space are uncorrelated; $\sigma_\text{QE-N}$ makes no such approximations. Finally, on the output side, the ``power spectrum method" returns a single error bar; the 'covariance method' provides a full bandpower covariance matrix.
\end{itemize}

 Of course, in reality not all delay modes are noise-dominated, and reliable error bars need to be placed even in signal-dominated regimes (whether this signal comes in the form of instrument systematics, foregrounds, or---ultimately---the cosmological signal). It is difficult to place rigorous error bars on bandpowers in these regimes: unless one has a physical model for all the systematics involved (with knowledge of their probability distributions), it is an ill-defined problem to ask how errors propagate. Unfortunately, the presence of unexplained (or at least not fully explained) systematics is the current state of affairs in $21\,\textrm{cm}$ cosmology, and truly rigorous error bars will need to wait for future work on the modeling of systematics.

Even with well-defined (if not perfectly characterized) systematics, the meaning of one's error bars is subtle. For instance, foregrounds such as a continuum of unresolved point sources can be appropriately treated as a random field. Given this, one's approach might be to say that the unresolved point sources contribute some effective power spectrum to the measurement. With such a formalism, there is a fundamental limit to how well these foregrounds can be characterized, since they come with their own form of cosmic variance. In other words, if one is trying to place constraints on foregrounds, one must account for the fact that the particular realization of foregrounds that we see may not be representative of foregrounds in general. This sort of error is difficult to compute in general, as the squared nature of the power spectrum means that the non-Gaussian---and therefore non-trivial---four-point function of the foregrounds needs to be known.

A goal of characterizing the general statistical properties of all possible foregrounds, however, may be unnecessarily ambitious. In particular, for a cosmological measurement one is not particularly concerned with the behaviour of a ``typical" foreground; one is primarily concerned with how our particular realization of foregrounds affect our observations. As a concrete example, if our Galaxy's synchrotron emission happens to be anomalously bright compared to a typical galaxy's synchrotron emission, it is our own brighter foregrounds that we need to deal with! With such a mindset, it is more appropriate to consider all foregrounds as non-random components of our data. By this, we do not mean that the foregrounds need to be spatially or spectrally constant; rather, we mean that in hypothetical random draws for taking ensemble averages, the cosmological signal and the instrumental noise change with each new realization, but the foregrounds remain the same. If the foregrounds are not formally random, our error bars are the result of instrumental noise (and in principle cosmic variance of the cosmological signal, although this contribution is small for current upper limits).

It is important to stress, however, that even if our error bars are due to the randomness of instrumental noise, the resulting error bars are \emph{not} simply what one obtains from imagining a noise-only measurement and propagating the noise fluctuations through to a power spectrum. This is because the power spectrum is a squared statistic. Thus, in the squaring of a measurement that contains both noise and a (non-random) signal, there are signal-noise cross-terms to contend with. These terms are zero in expectation, but do not have non-zero \emph{variance}. This means that knowledge of the signal (whether from systematics or foregrounds) is needed to correctly account for instrumental noise errors in non-noise-dominated regimes. 

\begin{table*}[!htbp]
    \centering
    \begin{tabular}{p{4cm}|p{6cm}|p{6cm}}
    \toprule
     Error Bar Type & Pros & Cons \\  
     \midrule
     Bootstrap ($\sigma_\text{bs}$) & Easy to implement with minimal \textit{a priori} assumptions; can be useful as a reference statistics in diagnosis of systematics & Not strictly applicable in the presence of non-independent and non-statistically stationary data samples
     \\ \hline 
     Power spectra from differenced visibilities ($P_\text{diff}$) & Data product close to raw data  & Provides noise \emph{realizations} rather than direct error bars, resulting in considerable scatter
     \\ \hline
     Power spectrum method ($P_\text{N}$ and $P_\text{SN}$) & Accurately captures variances/error bars in noise-dominated regimes (both $P_\text{N}$ and $P_\text{SN}$) and signal-dominated regimes ($P_\text{SN}$) & Does not contain covariance information between different bandpowers; $P_\text{SN}$ requires non-negativity prior on the signal, which slightly inflates errors; downstream data weightings using $P_\text{SN}$ at risk of signal loss
     \\ \hline
     Covariance method ($\sigma_\text{QE-N}$ and $\sigma_\text{QE-SN}$) & Same accuracy as $P_\text{N}$ and $P_\text{SN}$ for variance information and additionally provides full covariance information  & Derivation assumes data is Gaussian; $\sigma_\text{QE-SN}$ requires non-negativity prior on the signal, which slightly inflates errors; downstream data weightings using $\sigma_\text{QE-SN}$ at risk of signal loss\\ 
 \hline
          Modified covariance method ($\tilde{\sigma}_\text{QE-SN}$) and modified power spectrum method $\tilde{P}_\text{SN}$ & Eliminates conservative double counting of noise in noisy estimates of the signal   & Occasional error predictions that are slightly smaller than instrumental noise expectations from $\sigma_\text{QE-N}$ and $P_\text{N}$\\ 
     \bottomrule
\end{tabular}
    \caption{A summary of the advantages and disadvantages of different error estimation methods in 21\,cm power spectrum estimation.}
    \label{tab:summary}
\end{table*}

\begin{itemize}
    \item In short, even if we lower our ambitions and forgo incorporating knowledge about signal \emph{statistics} into our error calculations, understanding the signal itself is necessary for computing noise-sourced error bars. This requirement is where noise-only computations like $P_\text{N}$ and $\sigma_\text{QE-N}$ fall short. 

\item This shortcoming is remedied by generalized versions of $P_\text{N}$ and $\sigma_\text{QE-N}$, which we dub $P_\text{SN}$ and $\sigma_\text{QE-SN}$. These are given by Equations \eqref{eq:P_SN} and \eqref{eq:QE-SN}. The key idea is that in signal dominated regimes, the measured data itself can be a good approximation to the signal. Thus, we may reinsert the data in an appropriate way to capture signal terms in our general expressions. Figures \ref{fig:toy_single_ps} and \ref{fig:toy_average_ps} show that these error bars work well in both signal-dominated and noise-dominated regimes.

\item Although we treat foregrounds and systematics as a single signal term that is directly estimated from measured data in this paper, we note that for future high-sensitivity detections, more elaborate modeling of both are needed. Of course, there is also the possibility of unknown systematic effects, which our formalism does not account for.

\item Moreover, two cautionary warnings are in order when applying Equations \eqref{eq:P_SN} and \eqref{eq:QE-SN}. The first is that because the measured data are now part of the error bars themselves, it can be dangerous to use these error bars to inform data weightings for downstream averages in one's pipeline (e.g., in further incoherent time averaging of power spectra or in incoherent averaging of power spectra from different baselines). If the data weightings are coupled to the data themselves, our so-called quadratic estimators are no longer quadratic. As shown in \citet{cheng2018characterizing}, a blind application of the usual methods for normalizing quadratic estimators leads to power spectrum estimates that are biased low (``signal loss"). For this reason, while $P_\text{SN}$ and $\sigma_\text{QE-SN}$ are fine ways to compute error bars, we recommend that any error-motivated data weightings be based on $P_\text{N}$ and $\sigma_\text{QE-N}$ instead.

\item The second warning is that there almost certainly exist regimes that are neither signal- nor noise-dominated, where signal and noise are comparable in magnitude. Here, it becomes necessary to contend with the fact that a noisy measurement of the signal can be unphysically negative. Said differently, if our estimate of the signal itself contains noise, we are in effect double counting the noise in our error computations. One approach is to enact a hard prior on the positivity of the signal. This is what was done in all computations of $P_\text{SN}$ and $\sigma_\text{QE-SN}$ in this paper. However, Figures \ref{fig:toy_single_ps} and \ref{fig:toy_average_ps} show that this has the effect of inflating the error bars. Given that this is a conservative bias on the errors, this may or may not be appropriate depending on one's application.

\item A slightly more accurate approach is to assume that instrumental noise is Gaussian distributed and to quantitatively predict and correct the noise bias in the errors. Implementing this correction gives $\tilde{P}_\text{SN}$ and $\tilde{\sigma}_\text{QE-SN}$, which are given by Equations~\eqref{eq:modified_P_SN} and \eqref{eq:modified_QE-SN} respectively. Figures \ref{fig:toy_single_ps} and \ref{fig:toy_average_ps} show that this corrects the bias and gives error bars that are no longer overly conservative. However, because this correction is designed to give unbiased errors \emph{in expectation}, it will occasionally give error bars that are slightly smaller than the error predicted by noise-only estimators such as $P_\text{N}$. In practice, however, we find that this is a reasonably rare occurence.
\end{itemize}

With the aforementioned difficulties with error estimation in the presence of poorly characterized signals, one may be tempted to make use of more empirically based error estimates. These estimates also come with their strengths and weaknesses:
\begin{itemize}
    \item As discussed in Section \ref{subsec:P_diff}, $P_\text{diff}$ from frequency-differenced data suffers from a bias at low delays. Figure \ref{fig:freq_difference} shows that even at reasonably high delays $\sim 1500$ ns, the bias can be significant. Thus, while $P_\text{diff}$ from  frequency-differenced data is a useful asymptotic check at high delays, it is not a robust estimator of our errors. Implementing $P_\text{diff}$ using time-differenced data does not have the delay-dependent bias, as one can also see in Figure \ref{fig:freq_difference}. However, care must be taken to ensure that the time differencing is small enough to suppress any sky signal that is coherent between adjacent time samples \citep{dillon2015empirical}. In addition, with a differencing scheme one is ultimately constructing noise \emph{realizations}, not noise statistics. The resulting error bars thus show considerable scatter. In that sense, the analytically propagated error bars vary in a more physically plausible---smoother---way with time and frequency.
    
    \item The problem of a noisy error bar estimate persists with $\sigma_\text{bs}$. However, bootstrapping has several appealing features that makes it a crucial check on the analytically propagated error bars. First, no assumptions are made regarding Gaussianity of the input data. Thus, the fact that our $\sigma_\text{bs}$ agree with our analytically propagated errors---which assumed the input noise in the visibilities---is an essential validation of our assumptions. In a similar way, $\sigma_\text{bs}$ may potentially capture increased variance due to systematics since it is a measure of uncertainties of total sky emission. 
    However, the bootstrap method is known to suffer from some important limitations. For example, as noted in Appendix \ref{ap:bootstrap}, if systematics are correlated between samples, the bootstrap method tends to underestimate errors. Also, bootstrapped error bars will be inflated from non-stationary effects such as sky brightness changes and non-redundancies between nominally identical baselines. Precisely how these non-stationary effects should be folded into one's error estimation is reserved for future work, but the correct approach will certainly be more sophisticated than a simple inflation of errors. That said, this increase in bootstrap errors due to non-stationarity can serve as a useful diagnostic for further examination of unexpected systematics.
\end{itemize}

In Table~\ref{tab:summary} we summarize the discussion in this section with an succinct listing of the pros and cons of each error estimation method.

\section{Conclusions}
\label{ref:conc}
%%%
In this paper, we have systematically studied a variety of error bar methodologies in 21\,cm power spectrum estimation. We have synthesized some of the common techniques in the literature, outlining their relative strengths and weaknesses in quantifying noise levels and in accounting for residual systematics.
Specifically, we considered a variety of types of error estimators, including
\begin{itemize}
\item Power spectrum methods. This includes the standard $P_\text{N}$ estimator for the noise power spectrum found in the literature \citep{ZFH2004,parsons2012sensitivity, pober2013, cheng2018characterizing, kern2020abscal} and the $P_\text{SN}$ estimator that involves cross products with signal power spectrum $P_\text{S}$, as detailed in \citet{kolopanis2019simplified}. Here we set $P_\text{S}$ to be the real values of experimentally observed power spectrum, which is a good approximation when the signal dominates the noise. Our implementation of $P_\text{SN}$ leads to a double-counting bias compared to $P_\text{N}$ which is considerable in noise-dominated regimes, and we show how a modified form $\tilde{P}_\text{SN}$ can eliminate this bias. 

\item Covariance methods. This consists of propagating a data covariance matrix between frequencies per timestamp and per baseline-pair through the quadratic estimator (QE) formalism to the bandpower covariance matrix \citep{liu2011method, dillon2014overcoming, liu2014epoch, liu2014epochparttwo}, including error metrics described here: $\sigma_\text{QE-N}$ for noise-dominated spectra and $\sigma_\text{QE-SN}$ that include signal-noise terms. These have identical \emph{variance} predictions as $P_\text{N}$ and $P_\text{SN}$ by construction but also provide bandpower covariance information.

\item Other methods. Other methods studied in this work includes the bootstrapping method that can lead to misreported errors when not handled carefully \citep{cheng2018characterizing}, as well as the method of using differenced visibilities as noise realizations propagated through a power spectrum estimator. We show that differencing in frequency is ill-advised for this approach. Differencing in time avoids some problems, but either differencing scheme generates error estimates that are rather scattered. However, we stress the importance of these more empirically based methods are useful cross-checks (e.g., in the manner performed in this paper) that can also be helpful diagnostics for systematics (e.g., \citealt{kolopanis2019simplified}).
\end{itemize}

Using simulations and real HERA Phase I data, we show that these methods are generally in agreement with each other, demonstrating their robustness and their applicability to future delay power spectrum measurements from HERA. Importantly, we show that for bandpowers that are not completely dominated by noise, one needs to go beyond the standard thermal noise estimates and account for signal-noise cross terms in order to fully describe the uncertainty on the band power. In a series of Appendices, we also examine sources of skewness in probability distributions of measured power spectrum bandpowers (Appendices~\ref{sec:crosscorrskewness} and \ref{ap:skewness_intermediate}), derive exact expressions for the probability distributions of incoherently summed delay power spectra (Appendix~\ref{ap:pdf}), and examine whether common baselines in the cross multiplication of multiple baseline \emph{pairs} affects assumptions about error independence (Appendix~\ref{ap:bootstrap}). The insights gained in this paper regarding error estimation are applicable in $21\,\textrm{cm}$ cosmology beyond HERA. They provide a foundation upon which to develop rigorous error estimation methods which will prove to be key in unlocking the potential of the $21\,\textrm{cm}$ line as a powerful probe of our high redshift universe.

\newpage
\begin{acknowledgments}
This material is based upon work supported by the National Science Foundation under Grant Nos. 1636646 and 1836019 and institutional support from the HERA collaboration partners.
This research is funded in part by the Gordon and Betty Moore Foundation.
HERA is hosted by the South African Radio Astronomy Observatory, which is a facility of the National Research Foundation, an agency of the Department of Science and Innovation.
Parts of this research were supported by the Australian Research Council Centre of Excellence for All Sky Astrophysics in 3 Dimensions (ASTRO 3D), through project number CE170100013.
G.~Bernardi acknowledges funding from the INAF PRIN-SKA 2017 project 1.05.01.88.04 (FORECaST), support from the Ministero degli Affari Esteri della Cooperazione Internazionale - Direzione Generale per la Promozione del Sistema Paese Progetto di Grande Rilevanza ZA18GR02 and the National Research Foundation of South Africa (Grant Number 113121) as part of the ISARP RADIOSKY2020 Joint Research Scheme, from the Royal Society and the Newton Fund under grant NA150184 and from the National Research Foundation of South Africa (grant No. 103424).
P.~Bull acknowledges funding for part of this research from the European Research Council (ERC) under the European Union's Horizon 2020 research and innovation programme (Grant agreement No. 948764), and from STFC Grant ST/T000341/1.
J.S.~Dillon gratefully acknowledges the support of the NSF AAPF award \#1701536.
N.~Kern acknowledges support from the MIT Pappalardo fellowship.
A.~Liu acknowledges support from the New Frontiers in Research Fund Exploration grant program, the Canadian Institute for Advanced Research (CIFAR) Azrieli Global Scholars program, a Natural Sciences and Engineering Research Council of Canada (NSERC) Discovery Grant and a Discovery Launch Supplement, the Sloan Research Fellowship, and the William Dawson Scholarship at McGill.
\end{acknowledgments}

\appendix
\section{Skewness in power spectra estimated from multiple identical baselines}
\label{sec:crosscorrskewness}
In this Appendix, we consider a source of skewness in probability distributions of delay spectra. In particular, we consider the noise properties of power spectra formed from a set of identical (``redundant") baselines. We show that even if each baseline is measuring Gaussian random noise with mean zero, the resulting power spectra will exhibit some skewness. We emphasize, however, that this skewness vanishes if one additionally splits the data into two distinct set of time samples (e.g., even and odd time stamps) and estimates power spectra that are not only cross-baselines but also cross-times.

As a concrete example, suppose that on the $i$th copy of a particular baseline we measure $\tilde{x}_i \equiv c_i + i d_i$ after taking the delay transform, where $c_i$ and $d_i$ are independently Gaussian distributed random variables with variance $\sigma^2 /2$. This represents the behavior of $\tilde{x}_i$ at noise-dominated delays. If only two identical baselines were available, cross multiplying them to obtain a power spectrum would yield
\begin{equation}
\tilde{x}_1 \tilde{x}_2^* = (c_1 + i d_1) (c_2 - i d_2) = (c_1 c_2 + d_1 d_2) + i (d_1 c_2 - c_1 d_2).
\end{equation}
Consider the real part. Since $c_1$ and $c_2$ are independent random variables, $c_1c_2$ is a symmetric distribution about zero (and in fact is given by $K_0$, the zeroth modified Bessel function of the second kind). The same reasoning holds for the $d_1 d_2$ term. Since $\{c_i \}$ and $\{d_i \}$ are independent, it follows that $c_1c_2$ and $d_1 d_2$ are also independent. The result is that the probability distribution for $c_1 c_2 + d_1 d_2$ is given by the convolution of the distributions for the individual terms. With the two contributing distributions both symmetric about zero, their convolution inherits this property, and is in fact given by the Laplacian distribution discussed in Section~\ref{subsec:toymodel}.

The situation is different when we have more than two baselines. Taking all possible pairwise combinations (excluding the multiplication of a baseline with itself to eliminate noise bias), we obtain
\begin{equation}
\label{eq:threebls}
\textrm{Re}[\tilde{x}_1 \tilde{x}_2^* + \tilde{x}_1 \tilde{x}_3^* + \tilde{x}_2 \tilde{x}_3^* ] = (c_1 c_2 + c_1 c_3 + c_2 c_3) + (d_1 d_2 + d_1 d_3 + d_2 d_3),
\end{equation}
where we have grouped our result into two terms that can be considered separately because $\{c_i \}$ and $\{ d_i \}$ are independent. Consider the first term. It has zero mean:
\begin{equation}
\langle c_1 c_2 + c_1 c_3 + c_2 c_3 \rangle = \langle c_1\rangle \langle c_2\rangle + \langle c_1\rangle \langle c_3 \rangle+ \langle c_2\rangle \langle c_3 \rangle =0
\end{equation}
because the different $\{c_i \}$ are independent. However, the resulting distribution has a skewness to it, which can be seen by the fact that the third moment is non-zero:
\begin{eqnarray}
\langle (c_1 c_2 + c_1 c_3 + c_2 c_3)^3 \rangle &=& \langle c_2^3 c_1^3+c_3^3 c_1^3+3 c_2 c_3^2 c_1^3+3 c_2^2 c_3 c_1^3+3 c_2 c_3^3 c_1^2 +6 c_2^2 c_3^2 c_1^2+3 c_2^3 c_3 c_1^2+3 c_2^2 c_3^3 c_1+3 c_2^3 c_3^2 c_1+c_2^3 c_3^3 \rangle \nonumber\\
&=& 6 \langle c_2^2 c_3^2 c_1^2 \rangle = 6 \langle c^2_2 \rangle \langle c_3^2\rangle  \langle c_1^2 \rangle \neq 0
\end{eqnarray}
[Of course, in principle we should be taking the cube of Equation~\eqref{eq:threebls} in its entirety, not just the first term. However, the independence of $\{c_i \}$ and $\{d_i \}$ means we reach the same conclusion.] The non-zero third moment shown here arises because the three terms that make up the sum are correlated as a triplet, even though each pair has no average covariance. For instance, the covariance between $c_1 c_2$ and $c_1 c_3$ is
\begin{equation}
\langle c_1 c_2 c_1 c_3\rangle -\langle c_1 c_2 \rangle \langle c_1 c_3\rangle = \langle c_1^2 \rangle \langle c_2 \rangle \langle c_3 \rangle = 0.
\end{equation}
This implies that even though $c_1 c_2$, $c_1 c_3$, and $c_2 c_3$ are not independent, for the purposes of computing the variance of the final result, one obtains the same result even if one pretends that these contributions are independent. This result is explored in more detail in the first half of Appendix~\ref{ap:bootstrap}

To summarize, the different moments of the distribution provide different insights into power spectrum estimation with different baseline pair combinations. The mean of the distribution is zero, indicating that there is no bias (as one might expect for cross-correlation spectra). The variance turns out to be the same expression as if we had completely independent baseline pairs, so the noise averages down with the number of baseline pairs as one might naively have expected them to (without worrying about correlations). However, the skewness is non-zero. This complicates the interpretation of null tests that implicitly assume that the probability distributions of noise-dominated delays are symmetric.

Importantly, these considerations do not apply when we consider the imaginary part, which is given by
\begin{equation}
\textrm{Im}[\tilde{x}_1 \tilde{x}_2^* + \tilde{x}_1 \tilde{x}_3^* + \tilde{x}_2 \tilde{x}_3^* ] =c_2 d_1+c_3 d_1-c_1 d_2+c_3 d_2-c_1 d_3-c_2 d_3.
\end{equation}
This has a third moment given by $\langle (c_2 d_1+c_3 d_1-c_1 d_2+c_3 d_2-c_1 d_3-c_2 d_3)^3 \rangle$. To get terms that are non-zero under the expectation value, we require terms that contain \emph{squares} of the random variables when we multiply out the polynomial. For example, the first term $c_2 d_1$ must be multiplied onto $c_2 d_3$, because there is no other $c_2$ term in the expression to pair to. This gives us $c_2^2 d_1 d_3$. However, we now need to multiply this onto $d_1 d_3$, or we end up with a stray $d_1$ and a stray $d_3$. But none of the terms are the product of two $\{ d_i \}$, so no matter what terms we pair this up with, it will average to zero. This logic applies to any of the terms, so the distribution of the imaginary part will not be skewed. Because of this, statistical tests involving the imaginary part of a power spectrum estimator can be more easily interpreted using symmetric distributions.

Our result here has implications for how one should avoid the noise bias in power spectrum measurements. Two commonly used methods for doing so are to cross-multiply either different identical baselines together or different time stamps together. Here we have shown that employing only one of these will incur a skewness. (While our discussion above focused on cross multiplying different baselines, the same conclusions hold if we consider cross multiplying more than two groups in time---after all, the indices in our mathematical expressions can simply be considered timestamp indices instead of baseline indices.) However, if we perform cross-multiplications across both time and baseline axes, the skewness vanishes. To see this, imagine that we split our data into odd and even time samples, labeled with superscripts ``o" and ``e" respectively. Equation~\eqref{eq:threebls} then becomes
\begin{equation}
\textrm{Re}[\tilde{x}_1^e \tilde{x}_2^{o*} + \tilde{x}_1^e \tilde{x}_3^{o*} + \tilde{x}_2^e \tilde{x}_3^{o*} ] = (c_1^{e} c_2^{o} + c_1^{e} c_3^{o} + c_2^{e} c_3^{o}) + (d_1^{e} d_2^{o} + d_1^{e} d_3^{o} + d_2^{e} d_3^{o}),
\end{equation}
and cubing this expression as before to compute the third moment, one finds no non-zero terms after taking the ensemble average.

\section{Variance of averaged power spectra from dependent baseline-pair samples}
\label{ap:bootstrap}
In this Appendix, we consider the effect of having common baselines between different baseline \emph{pairs} used to form power spectra. Inside a redundant baseline group consisting of $N_\text{bl}$ different baselines, then we can construct up to $N_\text{blp} = \frac{1}{2}N_\text{bl}(N_\text{bl}-1)$ different baseline pairs and we can form a power spectrum using each pair. Consider the averaged power spectrum over these baseline pairs and the variance of this average. The form of the averaged power spectrum is
\begin{equation}
\overline{P} = \frac{\sum_{(p,q>p)} P_{pq}}{\frac{1}{2}N_\text{bl}(N_\text{bl}-1)} \,,
\end{equation}
where the sum is over all possible $(p, q)$ pairs of baselines. The variance of the averaged power spectrum does not simply go down with $N_\text{blp}^{-1}$ because the data being averaged together are not fully independent of each other. For example, $P_{12}$ and $P_{13}$ both carry information from baseline \#1. 

Let the signal be $\tilde{s} \equiv a + b i$, and $\tilde{n}_p \equiv c_p + d_p i$ and $\tilde{n}_q \equiv c_q + d_q i$ be the noise realizations in the $p$th and $q$th baselines. The signal $\tilde{s}$ is identical in each baseline, since we are assuming that we are combining data from identical (``redundant") baselines. The random variables $c_p, d_p, c_q, d_q ...$ are IID normal variables with variance $\sigma^2$. In the foreground-negligible regime, recall from Equation \eqref{eq:wedge} that the average power spectrum is given by
\begin{equation}
\overline{P} = \frac{\sum_{(p,q>p)} n_p^* n_q}{\frac{1}{2}N_\text{bl}(N_\text{bl}-1)} = \frac{\sum_{(p,q>p)} c_p c_q + d_p d_q}{\frac{1}{2}N_\text{bl}(N_\text{bl}-1)} + i \frac{\sum_{(p,q>p)} c_p d_q - c_q d_p}{\frac{1}{2}N_\text{bl}(N_\text{bl}-1)}\,.
\end{equation}
We notice 
\begin{align}
\textrm{Var} \left( \sum_{(p,q>p)} c_p c_q \right)=& \Bigg{\langle} \sum_{(p,q>p)} c_p c_q \sum_{(r,t>r)} c_r c_t \Bigg{\rangle} - \left[\Bigg{\langle} \sum_{(p,q>p)} c_p c_q\Bigg{\rangle}\right]^2  = \Bigg{\langle} \sum_{(p,q>p)} c_p c_q \sum_{(r,t>r)} c_r c_t \Bigg{\rangle} \notag \\
=& \sigma^4 \left[\sum_{(p,q>p,r,t>r)} (\delta_{pr}\delta_{qt} + \delta_{pt}\delta_{qr}) \right]= \frac{N_\text{bl}(N_\text{bl}-1)}{2} \sigma^4 \,,
\end{align}
which means that the variance in the real part of $\overline{P}$ is $\frac{4\sigma^4}{N_\text{bl}(N_\text{bl}-1)}$. For the imaginary part we compute
\begin{align}
\textrm{Var} \left(\sum_{(p,q>p)} c_p d_q - c_q d_p \right) =& \Bigg{\langle} \sum_{(p,q>p)} \{c_p d_q - c_q d_p\} \sum_{(r,t>r)} \{c_r d_t - c_t d_r\} \Bigg{\rangle} - \left[ \Bigg{\langle}\sum_{(p,q>p)} \{c_p d_q - c_q d_p\}\Bigg{\rangle} \right]^2  \notag \\
=& \Bigg{\langle} \sum_{(p,q>p)} \{c_p d_q - c_q d_p\} \sum_{(r,t>r)} \{c_r d_t - c_t d_r\} \Bigg{\rangle} = \sigma^4 \left[\sum_{(p,q>p,r,t>r)} (2\delta_{pr}\delta_{qt} - 2\delta_{pt}\delta_{qr})\right] \notag \\
=& N_\text{bl}(N_\text{bl}-1)\sigma^4 \,,
\end{align}
so that the variance of the imaginary part of $\overline{P}$ is also $\frac{4\sigma^4}{N_\text{bl}(N_\text{bl}-1)}$. Since the number of baseline pairs is given by $N_\text{bl} (N_\text{bl}-1)/2$ and $2 \sigma^4$ is the variance we would expect to get from a single baseline pair, we can see that $\overline{P}$ averages down in a manner that is identical to the scenario where the baseline pairs are independent.

In foreground-dominant regimes, the average power spectrum goes to 
\begin{equation}
\overline{P} = \frac{\sum_{(p,q>p)} s^*s + s^*n_q + n_p^*s}{\frac{1}{2}N_\text{bl}(N_\text{bl}-1)} = \frac{\sum_{(p,q>p)} a^2 + b^2 + a(c_p + c_q) + b(d_p + d_q)}{\frac{1}{2}N_\text{bl}(N_\text{bl}-1)} + i \frac{\sum_{(p,q>p)} a(d_q - d_p) + b(c_p - c_q)}{\frac{1}{2}N_\text{bl}(N_\text{bl}-1)}\,.
\end{equation}
The variance in the real part is $\frac{4(a^2+b^2)\sigma^2}{N_\text{bl}}$ and the variance in the imaginary part is $\frac{4(N_\text{bl}+1)(a^2+b^2)\sigma^2}{3N_\text{bl}(N_\text{bl}-1)}$. They now go down roughly as $N_\text{blp}^{-1/2}$ and are larger than the variance from independent samples by factors of $(N_\text{bl}-1)$ and $(N_\text{bl}+1)/3$ respectively.

\section{Time-Differenced Visibilities As Noise Estimators}
\label{ap:time_diff}
In this Appendix, we establish the validity of using time-differenced visibilities as a way to estimate noise error bars. The key idea is that if we form residuals of data vectors $x_p(\nu, t)$ by subtracting data from the $p$th baseline in adjacent time bins ($t_1$ and $t_2$) from each other, the result should be noise dominated. The same holds true for delay-transformed visibilities, where the residual can be written as $\tilde{n}_{p}(\tau, t_2) - \tilde{n}_p(\tau, t_1)$. Suppressing $\tau$ and demoting the time variable to a subscript for notational brevity, we write $\tilde{n}_{p,t} = c_{p,t} + d_{p,t} i$, where $c_p, d_p ...$ are IID normal variables with variance $\sigma^2$. The power spectra constructed from such residuals are 
\begin{align}
\label{eq:pnn}
    P_\text{diff} & =\frac{(\tilde{n}_{1,t2} - \tilde{n}_{1,t_1})^*}{\sqrt{2}}\frac{(\tilde{n}_{2,t2} - \tilde{n}_{2,t_1})}{\sqrt{2}} \notag \\
    & =   \left[ \frac{(c_{1,t2}-c_{1,t1})}{\sqrt{2}}\frac{(c_{2,t2}-c_{2,t1})}{\sqrt{2}} + \frac{(d_{1,t2}-d_{1,t1})}{\sqrt{2}}\frac{(d_{2,t2}-d_{2,t1})}{\sqrt{2}}  \right] \notag \\
    & \phantom{==} + \left[ \frac{(c_{1,t2}-c_{1,t1})}{\sqrt{2}}\frac{(d_{2,t2}-d_{2,t1})}{\sqrt{2}}-\frac{(c_{2,t2}-c_{2,t1})}{\sqrt{2}}\frac{(d_{1,t2}-d_{1,t1})}{\sqrt{2}}\right]i \,.
\end{align} 
From this, we see that
\begin{equation}
\bigg{\langle} \left[ \text{Re}(P_\text{diff})\right]^2 \bigg{\rangle} = \bigg{\langle} \left[\frac{(c_{1,t2}-c_{1,t1})}{\sqrt{2}}\frac{(c_{2,t2}-c_{2,t1})}{\sqrt{2}} + \frac{(d_{1,t2}-d_{1,t1})}{\sqrt{2}}\frac{(d_{2,t2}-d_{2,t1})}{\sqrt{2}} \right]^2 \bigg{\rangle} = \langle c_1^2\rangle\langle c_2^2\rangle + \langle d_1^2\rangle\langle d_2^2\rangle = 2\sigma^4.
\end{equation}
This is again the variance expected for a noise-dominated power spectrum. Therefore, what we have shown is that $|\text{Re}(P_\text{diff})|$ can serve as an estimator that \emph{in expectation} is equal to the correct noise errors for the measured power spectrum $P_{\tilde{x}_1\tilde{x}_2}$ in noise-dominated regimes. However, since this result only holds in expectation, we expect that in practice it will exhibit considerable scatter as an error estimate.

\section{Signal dependent error bar from power spectrum method}
\label{ap:fg_dependent_var}
In this Appendix we derive an expression for the variance on the power spectrum in the presence of foregrounds or systematics (or any ``signal"). A similar derivation is presented in \citet{kolopanis2019simplified}. Given two delay spectra $\tilde{x}_1 = \tilde{s} + \tilde{n}_1$ and $\tilde{x}_2 = \tilde{s} + \tilde{n}_2$, the power spectra formed from $\tilde{x}_1^* \tilde{x}_2$ is 
\begin{align}
P_{\tilde{x}_1\tilde{x}_2} & = \tilde{s}^*\tilde{s} + \tilde{s}^*\tilde{n}_2 + \tilde{n}_1^*\tilde{s} + 
\tilde{n}_1^*\tilde{n}_2 \notag \\
& = \left[ a^2+b^2 + a(c_1+c_2) + b(d_1+d_2) + c_1c_2 + d_1d_2\right]  + \left[a(d_2-d_1)+b(c_1-c_2)+d_2 c_1-d_1 c_2 \right] i \,,
\end{align}
where we have written $\tilde{s} = a + b i$, $\tilde{n}_1 = c_1 + d_1 i$ and $\tilde{n}_2 = c_2 + d_2 i$. 

Consistent with the rest of the paper, we assume that $a$ and $b$ are not random variables, so that $\langle s \rangle = s$. The true sky power spectrum is then given by $P_{\tilde{s}\tilde{s}} = a^2 +b^2$, and $c_1$, $d_1$, $c_2$ and $d_2$ in noise parts are IID random normal variables. We then have
\begin{align}
\label{eq:var_real_ps}
\text{Var} \left[\text{Re} (P_{\tilde{x}_1\tilde{x}_2}) \right] &= \text{Var} \left[a^2+b^2 + a(c_1+c_2) + b(d_1+d_2) + c_1c_2 + d_1d_2 \right] \notag \\
& = 2(a^2+b^2)\langle c_1^2\rangle + 2\langle c_1^2\rangle^2  = \sqrt{2}P_{\tilde{s}\tilde{s}}P_\text{N} + P_\text{N}^2  = \sqrt{2}\langle \text{Re} (P_{\tilde{x}_1\tilde{x}_2})\rangle P_\text{N} + P_\text{N}^2 = P_\text{SN}^2 \,.
\end{align}
In the above we have used the relation $\text{var} ( c_1c_2 + d_1d_2) = 2\langle c_1^2\rangle^2 = P_\text{N}^2$, where $P_\text{N}$ is the analytic noise power spectrum. We have also used $P_{\tilde{s}\tilde{s}} = \langle \text{Re} (P_{\tilde{x}_1\tilde{x}_2})\rangle$. This shows that $P_\text{SN}$ is a general form for error bars in the existence of foregrounds or systematics (or again, any ``signal").

\section{Covariance method}
\label{ap:covariance}
In this Appendix we provide more explicit derivations of the expressions quoted in Section~\ref{subsec:cov_Method} for the covariance method of error estimation.
\subsection{Variance}
If $\hat{P}_\alpha$ is a complex number representing a power spectrum estimate of the $\alpha$th bandpower, its real part and imaginary part are given by $\frac{1}{2}(\hat{P}_\alpha + \hat{P}_\alpha^*)$ and $\frac{1}{2i}(\hat{P}_\alpha - \hat{P}_\alpha^*)$ respectively. The variance in the real part of $\hat{P}_\alpha$ is
\begin{equation}
\label{eq:appendixrealvar}
    \frac{1}{4}\left\{ (\langle \hat{P}_\alpha \hat{P}_\alpha \rangle - \langle \hat{P}_\alpha \rangle\langle \hat{P}_\alpha \rangle) + 2(\langle \hat{P}_\alpha  \hat{P}_\alpha^* \rangle - \langle \hat{P}_\alpha \rangle\langle \hat{P}_\alpha ^*\rangle) 
        + (\langle \hat{P}_\alpha ^*  \hat{P}_\alpha ^*\rangle - \langle \hat{P}_\alpha^*\rangle\langle \hat{P}_\alpha^*\rangle) \right\}\,, 
\end{equation}
while the variance in the imaginary part of $\hat{P}_\alpha$ is
\begin{equation}
\label{eq:appendiximagvar}
      -\frac{1}{4}\left\{ (\langle \hat{P}_\alpha \hat{P}_\alpha \rangle - \langle \hat{P}_\alpha \rangle\langle \hat{P}_\alpha \rangle) - 2(\langle \hat{P}_\alpha  \hat{P}_\alpha^* \rangle - \langle \hat{P}_\alpha \rangle\langle \hat{P}_\alpha ^*\rangle) 
        + (\langle \hat{P}_\alpha ^*  \hat{P}_\alpha ^*\rangle - \langle \hat{P}_\alpha^*\rangle\langle \hat{P}_\alpha^*\rangle) \right\}\,.
\end{equation}

Recall that $\hat{P}_\alpha$ is defined as $\hat{P}_\alpha = \bm{x}_1^\dagger \bm{E}^{12,\alpha} \bm{x}_2 = \sum_{ij} \bm{x}_{1,i}^*\bm{E}^{12,\alpha}_{ij}\bm{x}_{2,j}$. We define three set of matrices containing the whole two-point correlation information for the complex estimator $\bm{C}^{12}$, $\bm{U}^{12}$ and $\bm{G}^{12}$, such that
\begin{align}
\bm{C}^{12}_{ij} \equiv  \langle \bm{x}_{1,i} \bm{x}_{2,j}^* \rangle; \qquad \bm{U}^{12}_{ij} \equiv  \langle \bm{x}_{1,i} \bm{x}_{2,j}\rangle; \qquad \bm{G}^{12}_{ij} \equiv  \langle \bm{x}_{1,i}^* \bm{x}_{2,j}^*\rangle \,,
\end{align}
Equipped with these definitions, we can generate the following equations
\begin{align}
\langle \hat{P}_\alpha \hat{P}_\beta \rangle - \langle \hat{P}_\alpha \rangle\langle \hat{P}_\beta\rangle =& 
\sum_{ijkl}\langle \bm{x}_{1,i}^*\bm{E}^{12,\alpha}_{ij}\bm{x}_{2,j}\bm{x}_{1,k}^*\bm{E}^{12,\beta}_{kl}\bm{x}_{2,l}\rangle - \langle \bm{x}_{1,i}^*\bm{E}^{12,\alpha}_{ij}\bm{x}_{2,j}\rangle\langle \bm{x}_{1,k}^*\bm{E}^{12,\beta}_{kl}\bm{x}_{2,l}\rangle\notag\\
=&\sum_{ijkl}\bm{E}^{12,\alpha}_{ij}\bm{E}^{12,\beta}_{kl}(\langle \bm{x}_{1,i}^*\bm{x}_{2,j}\bm{x}_{1,k}^*\bm{x}_{2,l}\rangle - \langle \bm{x}_{1,i}^*\bm{x}_{2,j}\rangle\langle \bm{x}_{1,k}^*\bm{x}_{2,l}\rangle)\notag\\
=&\sum_{ijkl}\bm{E}^{12,\alpha}_{ij}\bm{E}^{12,\beta}_{kl}(\langle \bm{x}_{1,i}^*\bm{x}_{1,k}^*\rangle\langle \bm{x}_{2,j}\bm{x}_{2,l}\rangle + \langle \bm{x}_{1,i}^*\bm{x}_{2,l}\rangle \langle \bm{x}_{1,k}^*\bm{x}_{2,j}\rangle)\notag\\
=&\sum_{ijkl}\bm{E}^{12,\alpha}_{ij}\bm{E}^{12,\beta}_{kl} (\bm{G}_{ik}^{11}\bm{U}_{jl}^{22} + \bm{C}^{21}_{li}\bm{C}^{21}_{jk})\notag\\
=&\sum_{ijkl}( \bm{E}^{12,\alpha}_{ij}\bm{U}_{jl}^{22}\bm{E}^{21,\beta*}_{lk} \bm{G}_{ki}^{11} + \bm{E}^{12,\alpha}_{ij}\bm{C}^{21}_{jk} \bm{E}^{12,\beta}_{kl} \bm{C}^{21}_{li} ) \notag\\
=&\text{tr}(\bm{E}^{12,\alpha} \bm{U}^{22} \bm{E}^{21,\beta*} \bm{G}^{11}+\bm{E}^{12,\alpha} \bm{C}^{21} \bm{E}^{12,\beta} \bm{C}^{21}) \,,
\end{align}

\begin{align}
\langle \hat{P}_\alpha \hat{P}_\beta^*\rangle - \langle \hat{P}_\alpha\rangle \langle \hat{P}_\beta^*\rangle =&
\sum_{ijkl}\langle \bm{x}_{1,i}^*\bm{E}^{12,\alpha}_{ij}\bm{x}_{2,j}\bm{x}_{1,k}\bm{E}^{12,\beta*}_{kl}\bm{x}_{2,l}^*\rangle - \langle \bm{x}_{1,i}^*\bm{E}^{12,\alpha}_{ij}\bm{x}_{2,j}\rangle\langle \bm{x}_{1,k}\bm{E}^{12,\beta*}_{kl}\bm{x}_{2,l}^*\rangle\notag\\
=&\sum_{ijkl} \bm{E}^{12,\alpha}_{ij}\bm{E}^{12,\beta*}_{kl}(\langle \bm{x}_{1,i}^*\bm{x}_{2,j}\bm{x}_{1,k}\bm{x}_{2,l}^*\rangle - \langle \bm{x}_{1,i}^*\bm{x}_{2,j}\rangle\langle \bm{x}_{1,k}\bm{x}_{2,l}^*\rangle)\notag\\
=&\sum_{ijkl}  \bm{E}^{12,\alpha}_{ij}\bm{E}^{12,\beta*}_{kl}(\langle \bm{x}_{1,i}^*\bm{x}_{2,l}^*\rangle \langle \bm{x}_{1,k}\bm{x}_{2,j}\rangle + \langle \bm{x}_{1,i}^*\bm{x}_{1,k}\rangle\langle \bm{x}_{2,j}\bm{x}_{2,l}^*\rangle)\notag\\
=&\sum_{ijkl} \bm{E}^{12,\alpha}_{ij}\bm{E}^{12,\beta*}_{kl} ( \bm{G}_{il}^{12}\bm{U}_{kj}^{12} + \bm{C}^{11}_{ki}\bm{C}^{22}_{jl})\notag\\
=&\sum_{ijkl}( \bm{E}^{12,\alpha}_{ij}\bm{U}_{jk}^{21}\bm{E}^{12,\beta*}_{kl} \bm{G}_{li}^{21} + \bm{E}^{12,\alpha}_{ij}\bm{C}^{22}_{jl} \bm{E}^{21,\beta}_{lk} \bm{C}^{11}_{ki} ) \notag\\
=&\text{tr}(\bm{E}^{12,\alpha} \bm{U}^{21} \bm{E}^{12,\beta *} \bm{G}^{21}+\bm{E}^{12,\alpha} \bm{C}^{22} \bm{E}^{21,\beta} \bm{C}^{11}) \,,
\end{align}
and 
\begin{align}
\langle \hat{P}_\alpha^* \hat{P}_\beta^*\rangle - \langle \hat{P}_\alpha^* \rangle\langle \hat{P}_\beta^*\rangle=&
\sum_{ijkl}\langle \bm{x}_{1,i}\bm{E}^{12,\alpha*}_{ij}\bm{x}_{2,j}^*\bm{x}_{1,k}\bm{E}^{12,\beta*}_{kl}\bm{x}_{2,l}^*\rangle - \langle \bm{x}_{1,i}\bm{E}^{12,\alpha*}_{ij}\bm{x}_{2,j}^*\rangle\langle \bm{x}_{1,k}\bm{E}^{12,\beta*}_{kl}\bm{x}_{2,l}^*\rangle \notag\\
=&\sum_{ijkl} \bm{E}^{12,\alpha*}_{ij}\bm{E}^{12,\beta*}_{kl}(\langle \bm{x}_{1,i}\bm{x}_{2,j}^* \bm{x}_{1,k}\bm{x}_{2,l}^*\rangle - \langle \bm{x}_{1,i}\bm{x}_{2,j}^*\rangle\langle \bm{x}_{1,k}\bm{x}_{2,l}^*\rangle)\notag\\
=&\sum_{ijkl}  \bm{E}^{12,\alpha*}_{ij}\bm{E}^{12,\beta*}_{kl}(\langle \bm{x}_{1,i}\bm{x}_{1,k}\rangle \langle \bm{x}_{2,j}^*\bm{x}_{2,l}^*\rangle + \langle \bm{x}_{1,i}\bm{x}_{2,l}^*\rangle\langle \bm{x}_{2,j}^*\bm{x}_{1,k}\rangle)\notag\\
=&\sum_{ijkl} \bm{E}^{12,\alpha*}_{ij}\bm{E}^{12,\beta*}_{kl} ( \bm{G}_{jl}^{22}\bm{U}_{ik}^{11} + \bm{C}^{12}_{il}\bm{C}^{12}_{kj})\notag\\
=&\sum_{ijkl}( \bm{E}^{21,\alpha}_{ji}\bm{U}_{ik}^{11}\bm{E}^{12,\beta*}_{kl} \bm{G}_{lj}^{22} + \bm{E}^{21,\alpha}_{ji}\bm{C}^{12}_{il} \bm{E}^{21,\beta}_{lk} \bm{C}^{12}_{kj} ) \notag\\
=&\text{tr}(\bm{E}^{21,\alpha} \bm{U}^{11} \bm{E}^{12,\beta*} \bm{G}^{22}+ \bm{E}^{21,\alpha} \bm{C}^{12} \bm{E}^{21,\beta} \bm{C}^{12})\,,
\end{align}
where $\bm{E}^{12,\alpha*}_{ij}= \bm{E}^{21,\alpha}_{ji}$. Setting $\alpha = \beta$ in these equations then allows us to evaluate Equations~\eqref{eq:appendixrealvar} and \eqref{eq:appendiximagvar}.

\subsection{Covariance}
The covariance between the real part of $\hat{P}_\alpha$ and the real part of $\hat{P}_\beta$ is 
\begin{equation}
    \frac{1}{4}\left\{ (\langle \hat{P}_\alpha \hat{P}_\beta \rangle - \langle \hat{P}_\alpha \rangle\langle \hat{P}_\beta \rangle) + (\langle \hat{P}_\alpha  \hat{P}_\beta^* \rangle - \langle \hat{P}_\alpha \rangle\langle \hat{P}_\beta ^*\rangle) + (\langle \hat{P}_\alpha^*  \hat{P}_\beta \rangle - \langle \hat{P}_\alpha^* \rangle\langle \hat{P}_\beta \rangle)
        + (\langle \hat{P}_\alpha ^*  \hat{P}_\beta ^*\rangle - \langle \hat{P}_\alpha^*\rangle\langle \hat{P}_\beta^*\rangle) \right\}\,,
\end{equation}
and the covariance between the imaginary part of $\hat{P}_\alpha$ and the imaginary part of $\hat{P}_\beta$ is 
\begin{equation}
    \frac{1}{4}\left\{ (\langle \hat{P}_\alpha \hat{P}_\beta \rangle - \langle \hat{P}_\alpha \rangle\langle \hat{P}_\beta \rangle) - (\langle \hat{P}_\alpha  \hat{P}_\beta^* \rangle - \langle \hat{P}_\alpha \rangle\langle \hat{P}_\beta ^*\rangle) - (\langle \hat{P}_\alpha^*  \hat{P}_\beta \rangle - \langle \hat{P}_\alpha^* \rangle\langle \hat{P}_\beta \rangle)
        + (\langle \hat{P}_\alpha ^*  \hat{P}_\beta ^*\rangle - \langle \hat{P}_\alpha^*\rangle\langle \hat{P}_\beta^*\rangle) \right\}\,.
\end{equation}
These can be evaluted in the same way as the variances above.

\section{Skewness in distributions of power spectra at intermediate delays}
\label{ap:skewness_intermediate}
In this Appendix, we consider the probability distribution functions of power spectra where neither signals (e.g., foregrounds) or noise are dominant and both must be considered. Using the same notation as Appendix \ref{ap:fg_dependent_var}, the power spectra formed from $\tilde{x}_1 = \tilde{s} + \tilde{n}_1$ and $\tilde{x}_2 = \tilde{s} + \tilde{n}_2$ is 
\begin{align}
P_{\tilde{x}_1\tilde{x}_2} & = \tilde{s}^*\tilde{s} + \tilde{s}^*\tilde{n}_2 + \tilde{n}_1^*\tilde{s} + 
\tilde{n}_1^*\tilde{n}_2 \notag \\
& = \left[ a^2+b^2 + a(c_1+c_2) + b(d_1+d_2) + c_1c_2 + d_1d_2 \right] + \left[ a(d_2-d_1)+b(c_1-c_2)+d_2 c_1-d_1 c_2\right]i \,.
\end{align}

Note that $a$ and $b$ are constants and $c_1$, $d_1$, $c_2$ and $d_2$ are IID randomly normal variables. For the real part of $P_{\tilde{x}_1\tilde{x}_2}$, we have
\begin{equation}
    \langle \text{Re} (P_{\tilde{x}_1\tilde{x}_2})\rangle = a^2+b^2\,.
\end{equation}
After subtracting from the mean, its third moment is 
\begin{align}
    \bigg{\langle} \left[\text{Re} (P_{\tilde{x}_1\tilde{x}_2})-(a^2+b^2)\right]^3 \bigg{\rangle} &= \bigg{\langle} \left[ a(c_1+c_2) + b(d_1+d_2) + c_1c_2 + d_1d_2\right]^3 \bigg{\rangle}= 6 \langle a^2c_1^2 c_2^2 + b^2 d_1^2 d_2^2 \rangle >0 \,.
\end{align}
This non-vanishing third moment implies that the probability distribution of power spectra is skewed. This skewness disappears for either signal-dominated or noise-dominated cases. These results are evident in the histograms shown in Figure~\ref{fig:toy_single_ps}.

\section{Probability distribution for an incoherent sum of delay transform-estimated power spectra}
\label{ap:pdf}
In this Appendix, we derive the probability distribution for noise in a power spectrum that has been formed by the incoherent (i.e., after squaring) averaging of power spectra from individual time integrations. The resulting probability distribution is used in Figures~\ref{fig:null_test_noise}, \ref{fig:inpainted_only}, and \ref{fig:null_test_IDR2.2} to validate our error bar methodology.

For a noise-dominated delay power spectrum estimate, the power spectrum value $u$ measured at one instant in time is distributed as a double exponential:
\begin{equation}
p(x) = \frac{1}{ \sigma \sqrt{2}} \exp\left( - \frac{\sqrt{2} |u|}{\sigma}\right),
\end{equation}
where it is assumed that the power spectra are estimated by cross-correlation---thus eliminating noise bias---and where $\sigma$ is the standard deviation on the resulting power spectrum.

Now suppose we average together a number of these power spectra. Let the power spectrum value at the $i$th time step be given by $u_i$. The average value is then
\begin{equation}
z \equiv \sum_i w_i u_i,
\end{equation}
where $\{ w_i \}$ are a set of weights. Note that the error on each $x_i$ may be different, so we define
\begin{equation}
p_i(u_i) = \frac{1}{ \sigma_i \sqrt{2}} \exp\left( - \frac{\sqrt{2} |u_i|}{\sigma_i}\right).
\end{equation}
We now write down the probability distribution $p_+ (z)$ for $z$. First we define $y_i \equiv w_i u_i$, such that
\begin{equation}
p_i (y_i ) = \frac{1}{w_i \sigma_i \sqrt{2}} \exp\left( - \frac{\sqrt{2} |y_i|}{w_i \sigma_i}\right).
\end{equation}
With this notation, $z = \sum_i y_i$, and we can write down $z$ by using the fact that the probability distribution of a sum of two random variables is the convolution of their individual distributions. By the convolution theorem, this is equivalent to multiplying the Fourier transforms of the individual probability distributions $\widetilde{p}_i (k)$, and thus 
\begin{equation}
p_+ (z) =\int \frac{dk}{2\pi} e^{ikz} \prod_i \widetilde{p}_i (k)=\int \frac{dk}{2\pi} e^{ikz} \prod_i \frac{1}{1+ w_i^2 \sigma_i^2 k^2/2},
\end{equation}
where we have used the fact that in our case, $\widetilde{p}_i (k) = (1+ w_i^2 \sigma_i^2 k^2/2)^{-1}$.This integral can be evaluated by contour integration, giving
\begin{equation}
\label{eq:SumOfLaplace}
p_+ (z) = \sum_j \frac{e^{-|z| \sqrt{2} / w_j \sigma_j} }{w_j \sigma_j \sqrt{2} } \prod_{i\neq j} \frac{1}{1- w_i^2 \sigma_i^2 /w_j^2 \sigma_j^2 }.
\end{equation}
This is a weighted sum of double exponential distributions, and the curves in Figures~\ref{fig:null_test_noise}, \ref{fig:inpainted_only}, and \ref{fig:null_test_IDR2.2} labeled ``Sum of Laplacians" are plots of this formula.

In closing, we note one peculiarity about this derivation---our contour integration assumed that none of the $w_i \sigma_i$ values were exactly equal. In principle, this is a reasonable assumption, since for a drift scan telescope that is sky noise dominated the noise power is continually changing from one time integration to the next. In practice, however, if this change is happening slowly, two adjacent time integrations may have similar enough noise properties to make Equation \eqref{eq:SumOfLaplace} numerically problematic. If this is indeed the regime that one is in, it is advisable to instead use an approximate expression by letting $\sqrt{2} / w_i \sigma_i \equiv \kappa + \varepsilon_i$ and then Taylor expanding to leading order in $\varepsilon_i$.

\bibliographystyle{aasjournal}
\bibliography{bibtex}

\end{document}